\documentclass[12pt]{article}
\usepackage{epsfig, graphics, subfigure}
\usepackage{amsmath,amssymb,stackrel}
\usepackage[latin1]{inputenc}
\usepackage{graphicx}
\usepackage{epsfig}
\usepackage{amsfonts}
\usepackage{enumerate}
\usepackage{mathtools, bm}
\usepackage{amssymb, bm}
\usepackage[latin1]{inputenc}

\def\bkR{{\rm I\kern-.17em R}}

\def \1n{1\hskip -3pt \mbox{N}}

\def \Frac {\displaystyle \frac }
\def \Int {\displaystyle \int }
\def \Sum {\displaystyle \sum }

\begin{document}

\begin{titlepage}
\thispagestyle{empty}

\title{Uncertainty on the Reproduction Ratio in the SIR Model}
\author{S., ELLIOTT, $^{(1)}$ and C. GOURIEROUX $^{(2)}$ }
\date{December, 15, 2020\\ (Preliminary version)}
 \maketitle

\vspace{9cm}

The  authors gratefully acknowledge financial support from the ACPR chair "Regulation and Systemic Risk", and the Agence Nationale de la Recherche (ANR-COVID) grant ANR-17-EUR-0010.

\addtocounter{footnote}{1} \footnotetext{University of Toronto.}
\addtocounter{footnote}{1} \footnotetext{University of Toronto,  Toulouse School of Economics, and CREST.}
\end{titlepage}

\newpage
\begin{center}
Uncertainty on the Reproduction Ratio in the SIR Model\\

Abstract
\end{center}
\vspace{1em}

The aim of this paper is to understand the extreme variability on the estimated reproduction ratio $R_0$ observed in practice. For expository purpose we consider a discrete time stochastic version of the Susceptible-Infected-Recovered (SIR) model, and introduce different approximate maximum likelihood (AML) estimators of $R_0$. We carefully discuss the properties of these estimators and illustrate by a Monte-Carlo study the width of confidence intervals on $R_0$.\vspace{1em}

Keywords~: SIR Model, Reproduction Ratio, COVID-19, Approximate Maximum Likelihood, EpiEstim, Final Size.

\newpage
\section{Introduction}
\setcounter{equation}{0}\def\theequation{1.\arabic{equation}}

In the standard epidemiological model, the reproduction ratio\footnote{This terminology has been introduced in the epidemiological literature by McDonald (1952).}, that measures the expected number of persons that can be infected by a new infectious individual, plays a key role. Its value affects the explosive episode in the early phase of an epidemic, the size of the peak of infections as well as the epidemic final size [see e.g. Hethcote (2000), Ma, Earn (2006)]. It is followed daily, or weekly, as a simple indicator of either approaching, or receding from, the peak of an epidemic [see e.g. PHO (2020)], and is often used for sanitary policy. For instance, it may be used to fix the conditions of a partial lockdown, or to close the border to foreigners coming from other countries. ``Alert levels are frequently based on this new totemic figure" [Adam (2020)].

The reproduction ratio is a forward looking notion whose definition involves a conditional expectation. This is a model based notion that depends on the information and dynamic model used to evaluate the expectation. This ex-ante notion has to be distinguished from the ex-post analogue counting retrospectively the number of persons infected by a given individual.\footnote{The difference is similar to the difference between life expectancy and lifetime, or between volatility and realized volatility.} This model free ex-post notion cannot be computed without an accurate tracing process and is not immediately useful in a prediction perspective. \footnote{See however White, Pagano (2008) for an application with well-documented influenza on two troop ships in late fall of 1918.}

In practice this ratio is approximated which generates a large uncertainty regarding its value [see Sanchez, Blauer (1997), the discussion and Table 2 in Obadia et al. (2012), webFigure 10 in Cori et al. (2013)]. For instance the first estimates for COVID-19 in Wuhan, China, were between 1.9 and 6.4 [see Li et al. (2020), Riou, Althaus (2020), Sanche et al. (2020), Wu et al. (2020)]. It is so important that ``to calculate the official ratio of the United Kingdom, about ten groups present the results of their models to a dedicated government committee, which reaches consensus on a possible range. The individual models are not released" [Adam (2020)]. This uncertainty is due to the different interpretations and definitions of this ratio in models that underlie the estimation methods, to the estimation methods themselves [see Obadia et al. (2012), Cori et al. (2013) for standard estimation packages], and to the way they are adjusted to be applied in a rolling way  [Wallinga, Teunis (2004), Cori et al. (2013)]. Moreover the estimates are generally provided without confidence bands, whereas these bands can be large, especially in the early phases of an epidemic, and, at the limit, these estimates can be non consistent of the reproductive ratio of interest even if applied to a large population. \vspace{1em}

The aim of this paper is to analyze precisely the uncertainty and lack of robustness of the estimated reproduction ratios. For expository purpose, we focus on the standard Susceptible-Infected-Recovered (SIR) initially introduced by Kermack, McKendrick (1927) and largely used in the literature. This model is used to define without ambiguity the reproductive ratio.

In Section 2, we introduce a discrete time stochastic version of the SIR model, discuss the possibility to aggregate the individual medical histories without loss of information. We also rigorously define the notions of reproduction ratios and how they evolve during the epidemic. Statistical inference of SIR model is the topic of Section 3. Since the binomial distributions that underlie the SIR model can be approximated by either Poisson, or Gaussian distributions depending on the structure of the population and on the transition probabilities, different approximate maximum likelihood estimators of the ratios are  considered. They do not provide the same estimated values, nor do they have the same distributions when we perform the estimations in a Gaussian asymptotic framework. They can even be inconsistent in a Poisson asymptotic framework. This leads to Section 4 that contains a Monte-Carlo study to find confidence intervals valid for the different estimators and designs. The matrix variate definition of the reproductive number is introduced in Section 5 for a SIR model with heterogeneity. This leads to the introduction of within and between compartments reproductive ratios. Section 6 discusses an alternative definition of reproductive number, called instantaneous reproductive number introduced by Fraser (2007) which is based on a renewal equation for the evolution of infected individuals. This notion is the basis of a Bayesian estimation approach of the reproductive ratio, diffused by the EpiEstim R-package [Cori et al. (2013)]. The EpiEstim estimator is usually computed in a rolling way, but has to provide reasonable results in the standard SIR model. We discuss precisely why this approach considers a parameter of interest that does not correspond to the initial definition of the reproductive ratio and illustrate this feature by a Monte-Carlo study. We also discuss an alternative approach of the same type based on autoregressions of counts of new infected individual. Section 7 concludes. Appendix 1 provides a review of the main properties of the continuous time deterministic model and its Euler time discretization. Proofs of some estimation results and additional Monte-Carlo results are given in Appendices.

\section{Model and Observations}
\setcounter{equation}{0}\def\theequation{2.\arabic{equation}}

We consider a discrete time stochastic version of the SIR model, with three states : S=1 susceptible, I=2, infected, infectious, R=3 recovered, immunized (or removed). We also discuss the aggregation of observations, and the notion of the reproductive ratio.

\subsection{The model of individual histories}

The model specifies the joint distribution of individual medical histories. For each individual $i, (i=1,\ldots, n)$, and date $t, (t=0,1,\ldots,T)$, the variable $Y_{it}$ provides the state $j=1,2,3$ of individual $i$ at date $t$.\vspace{1em}

\textbf{Assumption A.1~:} The individual histories $[Y_{i,t}, t=0,1,\ldots, T],$

$i=1,\ldots,n$ are such that~:\\

i) the variables $Y_{i,t}, i=1,\ldots, n$ are independent conditional on past histories~:

$$
\underline{Y_{t-1}} = \left(\left[ Y_{i,t-1}, Y_{i,t-2}, \ldots, Y_{i,0}\right), i=1,\ldots, n\right].
$$

ii) They have the same transition matrix :

$$
P_t = (p_{jk} (t)),
$$

\noindent where $p_{jk} (t)$ is the probability to migrate from state $j$ at date $t-1$ to state $k$ at date $t$, conditional on the past.\vspace{1em}

iii) The structure of the transition matrix is~:

$$
P_t =
 \left(\begin{array}{ccc}
1-a N_2 (t-1)/n & aN_2 (t-1)/n & 0 \\ \\
0 & 1-c & c \\ \\
0 & 0 & 1
\end{array}
\right),
$$

\noindent where $N_2 (t-1)$ is the number of individuals in state $I=2$ at date $t-1$ and $a,c $ are parameters such that $a>0$, $0 < c < 1.$ The structure of the transition matrix characterizes the SIR model:

\begin{enumerate}[i)]
\item The last row of the matrix means that state $R=3$ is an absorbing state implying an individual cannot be infected twice.

\item The zero in the second row means that, after infection, the individual recovers, is immunized, and then cannot become at risk.

\item The zero in the first row means that the individual cannot recover without being infected first.

\item Parameter $c$ is constant and represents the intensity of recovering.

\item Parameter $a$ measures the contagion effect, and the intensity of being infected for an individual at risk is proportional to the proportion of infectious people
\end{enumerate}

Under Assumption A.1, we  deduce the joint distribution of $Y_{i,t}, i=1, \ldots, n,$

\noindent $ t=1,\ldots, T$ given the initial conditions $Y_{i,0}, i=1,\ldots,n$. Nothing is said about the initial drawing of the $Y_{i,0}, i=1,\ldots, n$. This conditional joint distribution is parameterized by two parameters $a$ and $c$, that are assumed to be independent of both $n,T$.

\subsection{Aggregated counts}

Under Assumption A.1, it is possible to aggregate the individual data without loosing information on parameters $a$ and $c$. We denote:

\begin{itemize}
\item $N_{jk} (t),j,k=1,2,3,$ the number of individuals transitioning from $j$ to $k$ between $t-1$ and $t$

\item $N_j (t), j=1,2,3,$ the number of individuals in state $j$ at date $t$

\item $\hat{p}_{jk} (t) = N_{jk} (t)/N_j (t-1),$ the sample analogue of $p_{jk} (t)$

\item $\hat{p}_j (t) = N_j (t)/n,$ the proportion of individuals in state $j$ at date $t$
\end{itemize}

It is known that the set of aggregates $\{ N_{jk} (t), j, k=1,2,3, t=1,\ldots, T\}$ is a sufficient statistic for the analysis (see Appendix 2). Therefore the analysis can be based on these aggregates only. In the SIR framework, these aggregates are related as shown in Table 1.
\newpage
\begin{center}
  Table 1~: The Aggregate Counts\vspace{1em}

  \begin{tabular}{|c|c|c|c|c|}
    \hline
         & 1 & 2 & 3 & Total \\ \hline
    1 & $N_{11} (t)$ & $N_{12} (t)$ & 0 & $N_1 (t-1)$ \\
    2 & 0 & $N_{22} (t)$ & $N_{23} (t)$ & $N_2 (t-1)$ \\
    3 & 0 & 0 & $N_{33} (t)$ & $N_3 (t-1)$ \\ \hline
    Total & $N_1 (t)$ & $N_2 (t)$ & $N_3 (t)$ & n \\
    \hline
  \end{tabular}
\end{center}

In particular,  the following relationships provide the cross-sectional counts in terms of the transition counts:\vspace{1em}

$$
\begin{array}{lcl}
N_1(t)   & = & N_{11}(t), \\
N_2(t)   & = & N_{12}(t) +  N_{22}(t),\\
N_3(t)   & = & N_{23}(t) +  N_{33}(t), \\
N_1(t-1) & = & N_{11}(t) +  N_{12}(t), \\
N_2(t-1) & = & N_{22}(t) +  N_{23}(t),\\
N_3(t-1) & = & N_{33}(t).
\end{array}
$$

For the SIR model, these equations can be solved to get the transition counts in terms of marginal counts. We have~:

$$
\begin{array}{lcl}
N_{11} (t) & = & N_1 (t), \\
N_{12} (t) & = & N_1 (t-1)-N_1 (t) = -\Delta N_1 (t), \\
N_{22} (t) & = & N_2 (t) + \Delta N_1 (t), \\
N_{23} (t) & = & N_2 (t-1) - N_2 (t) - \Delta N_1 (t) = -\Delta N_1 (t) -\Delta N_2 (t) = \Delta N_3 (t), \\
N_{33} (t) & = & N_3 (t-1),
\end{array}
$$

\noindent where $\Delta = Id-L$ is the difference operator.\vspace{1em}

We deduce the following result :\vspace{1em}

\textbf{Proposition 1~:} For the SIR model of Assumption A.1, the sequence $N(t) = [N_1 (t), N_2 (t), N_3 (t)]', t=0,\ldots, T,$ is also a sufficient statistic. Moreover the process $[N(t)]$ is an homogeneous Markov process.\vspace{1em}

Thus we have the same information in the transition counts and in the cross-sectional counts. This property is  not satisfied in other epidemiological models.

\subsection{Reproductive ratio}

Other summaries of the development of a disease have been introduced in the epidemiological literature. An important concept is the reproductive (or reproduction) ratio (number). It is defined by computing the expected number of individuals at risk that a new infected individual will infect during his/her infectious period. In our framework with constant recovery intensity the length of the infection/infectious period is stochastic, with a geometric distribution with elementary probability : $P (X = x) = c (1-c)^{x-1},$ survivor function : $P [X \geq x] = (1-c)^{x-1}, x=1,2,\ldots,$ and expectation : $EX = 1/c.$

We deduce the expected number of individuals infected by this individual newly infected at date $t$ as [Farrington, Whitaker (2003)]~:

\begin{eqnarray}
R^*_{0,t} & = & \Frac{a}{n} \Sum^\infty_{x=1} \left\{ E_t [N_1 (t+x-1)] (1-c)^{x-1} \right\} \nonumber \\
&=& \Frac{a}{n} \Sum^\infty_{x=0} \left\{ E_t [N_1 (t+x)] (1-c)^x]\right..
\end{eqnarray}

This expectation depends on the transmission rate $a$, of the survival function of the infectious period, but also of the expected proportion of people at risk. For instance, if the population at risk disappears : $N_1 (t) \simeq 0$, then $R_{0,t} = 0$ too. To adjust for the size of the population at risk and the medical notion of transmission, it is usually proposed to consider also~:

\begin{equation}
  R_{0,t} = \Frac{a}{N_1 (t)} \Sum^\infty_{x=0} [E_t [N_1 (t+x)] (1-c)^x].
\end{equation}

These quantities are called basic reproductive and effective reproductive numbers for $R_{0,t}$ and $R^*_{0,t}$, respectively. Under Assumption A.1, the predictions $E_t N_1 (t+x) = g [a,c, N_1 (t), N_2(t), N_3(t)]$ by the homogeneous Markov property, where $g$ is a nonlinear function independent of time. Therefore $R_{0,t}, R^*_{0,t}$ also depend on time through the marginal counts at time $t$.

In the literature, this time dependence is often disregarded by focusing at the very early phase (outbreak) of the epidemics. [see e.g. Hethcote (2000)].\vspace{1em}

At this date $t=0$, it is assumed that~:\vspace{1em}

i) $N_1(0) = n-\varepsilon, N_2 (0) = \varepsilon, N_3 (0) = 0$, where $\varepsilon, \varepsilon > 0,$ is very small.\vspace{1em}

This $\varepsilon$ corresponds to the first infected individuals, or the first cluster. Without this initial infection, the disease cannot appear in the population. In other words, the SIR model assumes a ``closed economy", except at the initial date.\vspace{1em}

ii) During the following days $N_1 (t) = n-\varepsilon (t),$ where $\varepsilon (t)$ is also small. An approximate formula for reproductive ratios is~:

\begin{equation}
  R_{0,0} = R^*_{0,0} \simeq a \Sum^\infty_{x=0} (1-c)^x = \Frac{a}{c},
\end{equation}

\noindent that is, the transmission rate times the expected length of the infection episode. This common value is called the initial reproductive ratio. However, during the epidemic, these measures can differ significantly.

\subsection{Simulation}

The conditional distributions of the count variables are easily deduced from Assumption A.1.\vspace{1em}

\textbf{Proposition 2~:} Under Assumption A.1,\vspace{1em}

i) $N_{12} (t)$ and $N_{23} (t)$ are independent given the past. $N_{12} (t)$ follows the binomial distribution $\mathcal{B} [N_1 (t-1), a \Frac{N_2 (t-1)}{n}].$ $N_{23} (t)$ follows the binomial distribution $\mathcal{B} [N_2 (t-1), c].$\vspace{1em}

ii) The process $[N_1 (t), N_2 (t)]$ is a Markov process. Its conditional distribution is obtained from the distribution of $[N_{12} (t), N_{23} (t)]$ by the change of variable~:

$$
\left\{
\begin{array}{lcl}
  N_1 (t) & = & N_1 (t-1) - N_{12} (t), \\ \\
  N_2 (t) & = & N_2 (t-1) + N_{12} (t) - N_{23} (t).
\end{array}
\right.
$$
\vspace{0.5em}

These results can be used to simulate the aggregate counts for given parameter values $a,c$ and given starting counts $N_1 (0), N_2 (0),$ $N_3 (0),$ along the following scheme, where $\stackrel{s}{\rightarrow}$ is a drawing in the binomial distributions of Proposition 1 i), and $\stackrel{d}{\rightarrow}$ the application of the deterministic relation in Proposition 1 ii).

\begin{center}
  Table 2~: Simulation Scheme\vspace{1em}

  \begin{tabular}{ccc}
      $[N_1 (0), N_2 (0), N_3 (0)]$ & $\stackrel{d}{\longrightarrow}$ & $[N_1 (1), N_2 (1), N_3 (1)] \stackrel{d}{\rightarrow}$ \\ \\
        $\downarrow s$ & $\nearrow d$ &  $\downarrow s$ \\ \\
    $[N_{12} (1), N_{23} (1)]$ &  &  \\ \\
      \end{tabular}
\end{center}

For simulations and by analogy with COVID-19, the parameter values can be fixed as~:

$c=0.07$, that corresponds to an expected infection\footnote{In the SIR model the infection and infectious periods are assumed the same. This is not the case for COVID-19.} period of approximately 14 days, $R_{0,0} = a/c$ between $0.5$ and $1.5$, that means $a$ between $0.095$ and $0.105$.\vspace{1em}

The initial structure for a population corresponding to the city of Toronto, say, can be $n= 3 000 000$ with a first cluster of $N_2 (0) = 50 \;[\mbox{with}\; N_3 (0)=0].$

Thus for $a \simeq 0.1$, at date $t=0, p_{12} (0) \simeq \Frac{0.1 \;\; 50}{3 000 000} = \Frac{1}{600 000}$.\vspace{1em}

Thus $p_{23} (t)$ is small and $p_{12} (0)$ very small as well as $p_{12} (t)$ at the beginning of the epidemic.\vspace{1em}

A simulated path is given in Figure 1. We observe the standard patterns~:

\begin{itemize}
\item A decreasing pattern for the size of the population at risk.

\item An increasing pattern for the number of immunized people.

\item The peak of the epidemic for the number of infected people, arising around one year in this simulation. The figure is given for a rather large number of days to highlight the asymptotic behaviour. For this SIR model, there is herd immunity [Allen (1994)], and the immunity ratio is around 55\%.
\end{itemize}

\newpage

\centerline{Figure ~1: Simulated Path}\vspace{1em}

\begin{flushleft}
\includegraphics[width=\textwidth]{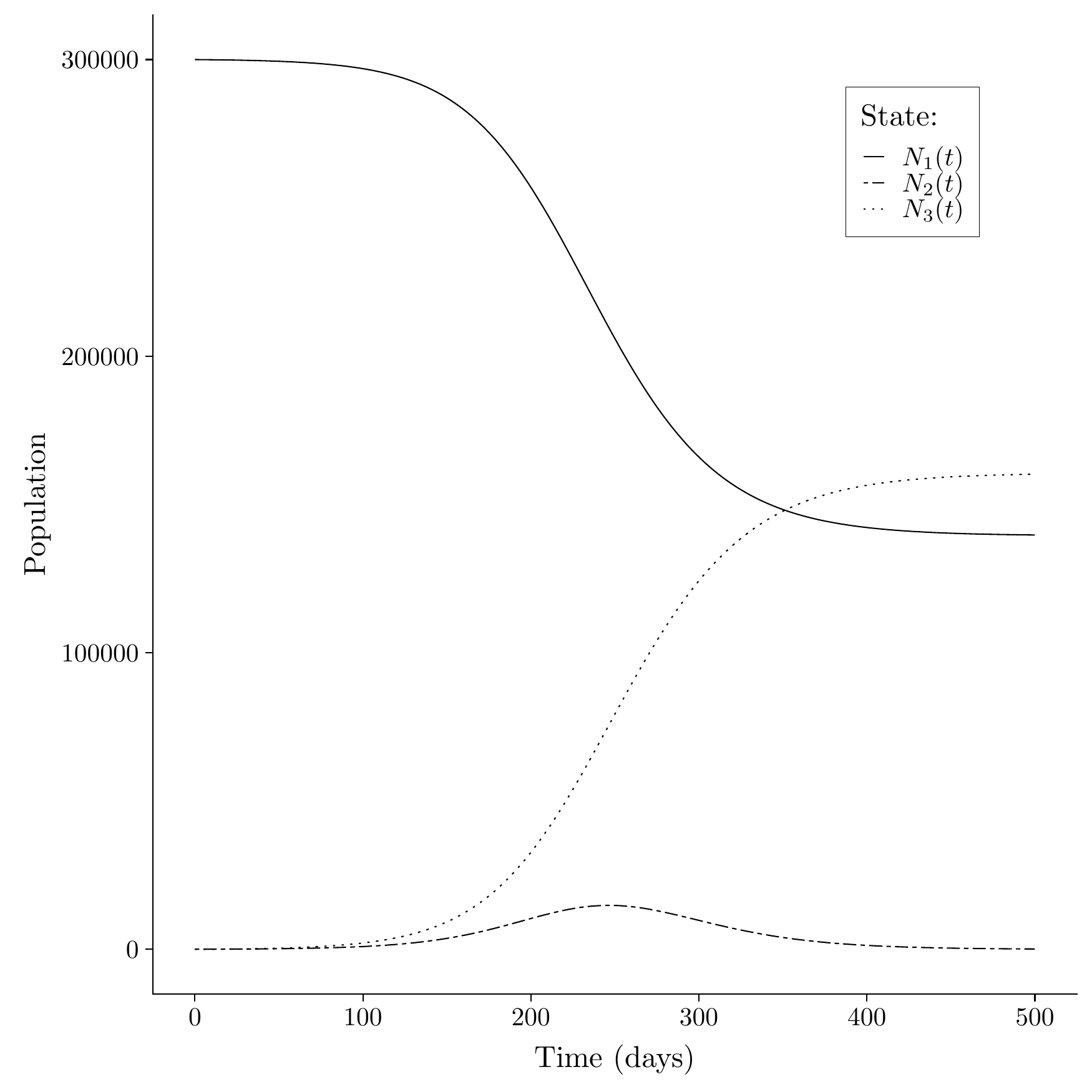}
\end{flushleft}

At each date  $t$, we can simulate and average of several future paths $N_1 (t+x), x=1,\ldots, 30,$ and compute the basic reproductive and effective reproductive numbers at $t$. These paths are reported in Figure 2 with a number of replications equal to $S = 100$. We observe that even the basic reproduction number, that is the number adjusted by the size of the population at risk, is not constant during the epidemics in the time discretized version of the SIR.
\newpage

\centerline{Figure 2~: Evolution of Basic and Effective Reproductive Ratios}\vspace{1em}

\begin{flushleft}
\includegraphics[width=\linewidth]{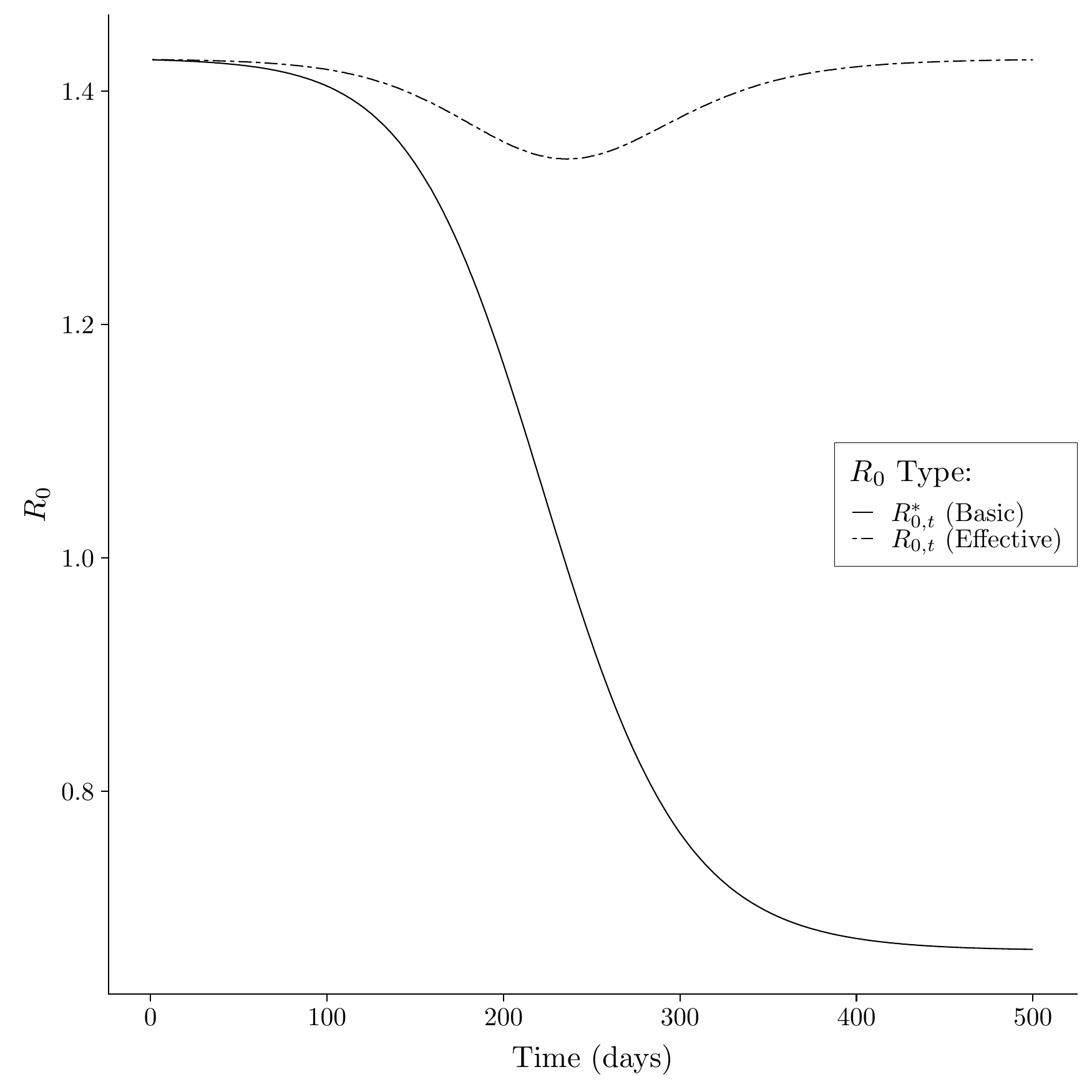}
\end{flushleft}

We also observe that the final level of the effective reproduction number is equal to its starting value. Indeed for large $t$, the size of the population at risk coincides with the final size and then $R_0 (\infty) = a/c$ too. The evolutions of Figure 2 are obtained with a length of 100 days for the future path of $N_1 (t).$ In practice the sum can be truncated and such a truncation can have an impact on the evaluation of $R_0$. Figure 3 provides the evolutions of reproduction ratios computed with 30, 60, and 100 days, respectively.\newpage

\centerline{Figure ~3: Evolution of Reproductive Ratio Under Truncation}\vspace{1em}
\begin{flushleft}
\includegraphics[width=\linewidth]{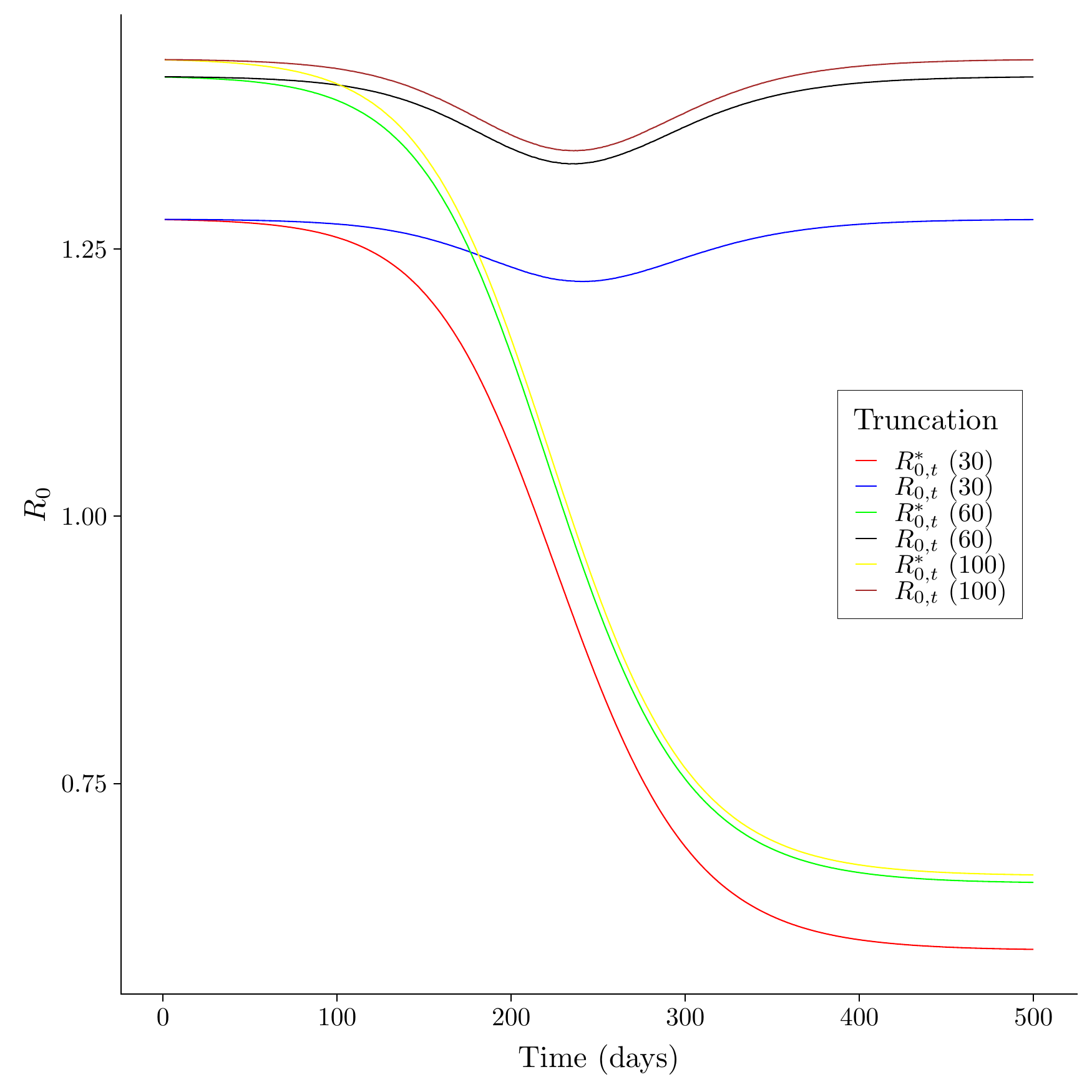}
\end{flushleft}

\newpage

\section{Estimation}
\setcounter{equation}{0}\def\theequation{3.\arabic{equation}}

\subsection{Challenges}

The estimation of a SIR model, and more generally of any epidemiological model, is a bit challenging for three main reasons.\vspace{1em}

i) The SIR model is a nonlinear dynamic model with chaotic properties [see e.g. Harko, Lobo, Mack (2014)]. This implies that small changes in parameter values $a,c$, in particular the estimation errors, can have a strong impact on the evolution of the process in the medium and long runs. It is known [see Allen (1994)] that the deterministic discrete time version of the SIR model satisfies herd immunity. However, in our stochastic framework the level of herd immunity, as well as the time at which it is reached, are very sensitive to the values of $a,c,$ and to the initial conditions.\vspace{1em}

ii) The evolution of the disease is non-stationary, as seen in Figure 1. If $R_{0,0} > 1$, the proportion of infected individuals increases up to a peak, then decreases towards an asymptotic stationary state. This non-stationarity makes it difficult to analyze the properties of the estimators as functions of the number of observation dates $T$. Moreover, $T$ is usually small, between 20-60 days, at the beginning of the epidemic.\vspace{1em}

iii) In contrast to the previous point, the cross-sectional dimension $n$ is very large and we expect an asymptotic theory when $n$ tends to infinity, $T$ being fixed. However, Proposition 2 shows the key role of the binomial distributions $\mathcal{B} [N_1 (t-1), p_{12} (t)]$ and $\mathcal{B} [N_2 (t-1), c], t=1,\ldots,T$. For an asymptotic analysis, what matters is not $n$, but rather the marginal counts $N_1 (t-1), N_2 (t-1)$. Whereas the susceptible population is often very large, at least at the beginning of the disease, the number of infected people is much smaller.\vspace{1em}

However, for large $N_1 (t-1), N_2 (t-1)$, we may apply the standard asymptotic results for a binomial distribution. That is, the possibility to approximate it by either a Poisson, or a Gaussian distribution. Thus, the approximation of $\mathcal{B} (N,p)$, say, is either $\mathcal{P} (Np),$  if $N\rightarrow \infty, p\rightarrow 0$, such that $Np \rightarrow \lambda > 0$, or $N[Np, Np (1-p)]$, if $N\rightarrow \infty$, $p$ being fixed. In our framework both $p_{23} (t) = c$ and $p_{12} (t)$ are small. The choice between the approximations depend on the magnitudes of $N_1 (t-1) p_{12} (t), N_2 (t-1) p_{23} (t)$, $t=1,\ldots, T,$. That is, the numbers of new infected and new recovered, respectively. \vspace{1em}

Loosely speaking, if they are smaller than 45-50, say, the Poisson approximation can be used, the Gaussian approximation, otherwise. But at the beginning of the epidemic and also at the end of the epidemic $N_{12} (t)$ and $N_{23}(t)$ are rather small, while being larger in the peak of the epidemic. Therefore, the approximation will depend on the observation date. They also depend on the size $n$ of the population of interest. For instance, this size is smaller if we want to consider a subpopulation of Toronto, say, males older than 75.\footnote{See also Zhang et al. (2020) for an analysis restricted to the analysis of the epidemic on the Diamond Princess cruise ship.}

\subsection{Mechanistic model}

A major part of the literature is based on a deterministic dynamic model, that assumes implicitly the possibility to closely approximate the theoretical transition probabilities by their frequency counterparts. That is, to use the Gaussian approximation.

More precisely, under Assumption A.1, we have~:

\begin{equation}
E_{t-1} \hat{p} (t) = P [\hat{p}_2 (t-1)]' \hat{p} (t-1).
\end{equation}

Therefore if $\hat{p} (t) \sim p(t)$, we get the following deterministic dynamic model for the $p(t)'s$:

\begin{equation}
  p(t) = P [p_2 (t-1)]' p(t-1).
\end{equation}

This is often called the mechanistic model [see Breto et al. (2009) and Appendix 1 for its link with the continuous time SIR model].

\subsection{(Approximate) Maximum Likelihood Estimator}

In our framework, the log-likelihood function $L(a,c)$ can be decomposed as a sum $L(a,c) = L_1 (a) + L_2 (c)$. This allows us to estimate separately $a$ and $c$ by focusing on the first and second rows of the (observed) transition matrix, respectively (see Appendix 2). Different log-likelihood functions can be considered such as, the true one based on the binomial distributions, or approximate ones based on either Poisson, or Gaussian approximations.

\subsubsection{Binomial log-likelihood}

We have~:

\footnotesize{\begin{eqnarray}
  L_1 (a) & = & \Sum^T_{t=1} \{ N_{11} (t) \log [1-a \hat{p}_2 (t-1)] + N_{12} (t) \log [a \hat{p}_2 (t-1)]\}, \\
  L_2 (c) & = & \Sum^T_{t=1} \{ N_{22} (t) \log (1-c)  + N_{23} (t) \log c\}.
\end{eqnarray}}
\normalsize

The ML estimator of $a$ is the solution of the first-order conditions~:

\begin{equation}
  - \Sum^T_{t=1} \left[
  \Frac{N_{11} (t) \hat{p}_2 (t-1)}{1-\hat{a} \hat{p}_2 (t-1)}\right] + \Frac{1}{\hat{a}} \Sum^T_{t=1} N_{12} (t) = 0,
\end{equation}

\noindent and has no closed form expression.\vspace{1em}

The ML estimator of $c$ is~:

\begin{eqnarray}
  \hat{c} & = & \Sum^T_{t=1} N_{23} (t) / \Sum^T_{t=1} N_2 (t-1) \nonumber \\
  &=& \Sum^T_{t=1} \left\{\Frac{N_2 (t-1)}{\Sum^T_{t=1} N_2 (t-1)} \hat{p}_{23} (t)\right\}.
\end{eqnarray}

This is a weighted combination of the dated transition frequencies.

\subsubsection{Poisson approximate log-likelihood}

We have~:

\small{
\begin{eqnarray}
  L^P_1 (a) & \propto & \Sum^T_{t=1} \{ N_{12} (t) \log [a N_1 (t-1) \hat{p}_2 (t-1)] - a N_1 (t-1) \hat{p}_2 (t-1)]\}, \\
  L^P_2 (c) & \propto & \Sum^T_{t=1} \{ N_{23} (t) \log [ N_2 (t-1)c] - N_2 (t-1) c]\}.
\end{eqnarray}}
\normalsize

We get Poisson approximate maximum likelihood (AML) estimators with closed form expressions~:

\begin{eqnarray}
  \hat{a}_P & = & n \Sum^T_{t=1} N_{12} (t) / \Sum^T_{t=1} [N_1 (t-1) N_2 (t-1)], \\
  \hat{c}_P & = &   \Sum^T_{t=1} N_{23} (t) / \Sum^T_{t=1} N_2 (t-1) = \hat{c}.
\end{eqnarray}

The first formula shows that $\hat{a}_P$ is a weighted average of the dated estimated transition coefficients~: $\hat{a}_t = N_{12} (t)/[N_1 (t-1) \hat{p}_2 (t-1)]$, with weights proportional to $N_1 (t-1) \hat{p}_2 (t-1).$

We deduce an analytical formula for the corresponding estimator of the initial reproductive number~:

\begin{equation}
  \hat{R}_{0,P} = \Frac{n \Sum^T_{t=1} N_{12} (t) \Sum^T_{t=1} N_2 (t-1)}{\Sum^T_{t=1} [N_1 (t-1) N_2 (t-1)] \Sum^T_{t=1} N_{12} (t)}.
\end{equation}

This formula can be used if $\Sum^T_{t=1} N_{23} (t)$ is non zero, that is if recovery has been observed.

\subsubsection{Gaussian approximate log-likelihood}

We have~:

\begin{eqnarray}
  L^G_1 (a) & \propto & -\Frac{1}{2} \Sum^T_{t=1} \log (a \hat{p}_2 (t-1) [1- a    \hat{p}_2 (t-1)]) - \Frac{1}{2} \Sum^T_{t=1} N_1 (t-1) \Frac{[\hat{p}_{12} (t) - a \hat{p}_2 (t-1)]^2}{a \hat{p}_2 (t-1) [1-a \hat{p}_2 (t-1)]}, \nonumber \\
  &&\\
  L^G_2 (c) & \propto & - \Frac{T}{2} \log [c(1-c)] - \Frac{1}{2} \Sum^T_{t=1} N_2 (t-1) \Frac{[\hat{p}_{23} (t) - c]^2}{c(1-c)}.
\end{eqnarray}

\subsubsection{Unfeasible Gaussian approximate log-likelihood}

The approximate log-likelihood is obtained by replacing the variance $a \hat{p}_2 (t-1) [1- a \hat{p}_2 (t-1)]$ by the estimate $\hat{p}_{12} (t) [1-\hat{p}_{12} (t)].$\footnote{This can be inconsistent when $n$ tends to infinity.}  We get~:

\begin{equation}
  L^{UG}_1 (a) = - \Frac{1}{2} \Sum^T_{t=1} \left\{N_1 (t-1) \Frac{(\hat{p}_{12}(t) - a \hat{p}_2 (t-1)]^2}{\hat{p}_{12} (t) (1-\hat{p}_{12} (t)]}\right\}.
\end{equation}

We get a closed form expression for $\hat{a}_{UG}$ that corresponds to an unfeasible Generalized Least Squares (GLS) estimator of a~:

\begin{equation}
  \hat{a}_{UG} = \Sum^T_{t=1} (N_1 (t-1) \hat{p}_2 (t-1)/ [1-\hat{p}_{12} (t))/\Sum^T_{t=1} \left[ \Frac{N_1 (t-1)\hat{p}_2 (t-1)^2}{\hat{p}_{12} (t) [1-\hat{p}_{12} (t)]}\right].
\end{equation}

\subsubsection{Poisson/Gaussian approximate log-likelihood}

When $n$ is large, $p$ small and $np$ large, the Poisson distribution $\mathcal{P}(np)$ can be approximated by a Gaussian distributions $N(np, np)$. Thus compared to the approximation in 3.3.3, the term in $p^2$ in the variance is disregarded. We have~:

\begin{align}
L^{PG}_1 (a) & \propto - \Frac{1}{2} \Sum^T_{t=1} \log [a \hat{p}_2 (t-1)] -\Frac{1}{2} \Sum^T_{t=1} \left\{N_1 (t-1)\Frac{[\hat{p}_{12} (t) - a \hat{p}_2 (t-1)]^2}{a\hat{p_2 (t-1)}}\right\} \\
L^{PG}_2 (c) &\propto  - \Frac{1}{2} T \log c - \Frac{1}{2} \Sum^T_{t=1} \left\{N_2 (t-1) \Frac{[\hat{p}_{23} (t) - c]^2}{c}\right\}
\end{align}

Then the AML estimates are positive solutions of polynomial equations of degree 2, that are~:

$$
\Frac{1}{T} \Sum^T_{t=1} \left\{N_1 (t-1) \hat{p}_2 (t-1)\right\} a^2 +
 a -  \Frac{1}{T} \Sum^T_{t=1} \left\{N_1 (t-1) \hat{p}_{12} (t) \right\}= 0,
$$

\noindent and

$$
\Frac{1}{T} \Sum^T_{t=1} N_2 (t-1) n c^2 + c -  \Frac{1}{T} \Sum^T_{t=1} \left\{N_2 (t-1) \hat{p}_{23} (t)\right\} = 0, \; \mbox{respectively.}
$$

To summarize, we get as many AML estimators of $a,c$ and of the initial reproduction number $R_{0,0} = a/c$ as (approximated) log-likelihoods. This can also explain the different approximations of $R_{0,0}$ published even when applied to the same series of aggregate counts.

\subsection{Properties of the AML estimators}

The properties of the AML estimators can be derived by Monte-Carlo as shown in Section 4. Their asymptotic properties depend on either the Poisson, or Gaussian asymptotics, depending on which is the most appropriate, and on the selected estimators. For instance, we may have chosen a Poisson AML estimator when the Gaussian asymptotic conditions were satisfied. In this case, whereas $B(N,p)$ is well approximated by $N[Np, Np(1-p)]$, it has been replaced by $\mathcal{P}(Np)$, which is close to $N(Np, Np)$. Therefore we have not used the right Gaussian approximation and have neglected the term in $p^2$.\vspace{1em}

For illustration we consider below two cases~: 

\begin{enumerate}[i)]
\item The behaviour of the Poisson AML estimator $\hat{a}_P$, when Poisson asymptotics are valid.

\item The behaviour of the binomial ML estimator $\hat{a}$, when Gaussian asymptotics are valid.
\end{enumerate}

\subsubsection{Poisson AML and Poisson asymptotics}

Let us consider the case $T=1$, that is two observations of the aggregates. The main results below will be valid for any finite $T$. Then we have~:

\begin{equation}
  \begin{array}{lcl}
  \hat{a}_P & = & n N_{12} (1)/N_1 (0) N_2 (0), \\ \\
  \hat{c}_P & = &  N_{23} (1)/N_2 (0), \\ \\
  \hat{R}_{0,P} & = & \hat{a}_P/\hat{c}_P = \Frac{N_{12} (1)}{N_{23}(1)} \Frac{n}{N_1 (0)}.
  \end{array}
\end{equation}

Conditional on $[N_1 (0), N_2 (0)], $ the estimators $\hat{a}_P$ and $\hat{c}_P$ are independent such that $\Frac{N_1 (0) N_2 (0)}{n} \hat{a}_P \sim \mathcal{P} [a \Frac{N_1 (0) N_2 (0)}{n}],$ $N_2 (0) \hat{c}_P \sim \mathcal{P} [c N_2 (0)]$.\vspace{1em}

We deduce that~: $E_0 \hat{a}_P = a, E_0 \hat{c}_P =c,$ that shows that the Poisson AML estimators are unbiased for $T=1$. Their variances are~:

\begin{equation}
  V_0 \hat{a}_P = \Frac{a n}{N_1 (0) N_2 (0)}, V_0 \hat{c}_P = \Frac{c}{N_2 (0)}.
\end{equation}

In practice $N_1 (0) \simeq n$, and $N_2 (0)$ is rather small $(<30$ or $40$, say) for Poisson asymptotics to be valid. Therefore both $V_0 (\hat{a}_P)$ and $V_0 (\hat{c}_P)$ are not small, even for large $n$ and we cannot expect the consistency of $\hat{a}_P, \hat{c}_P$ for $n$ large under Poisson asymptotics.\vspace{1em}

Moreover, at the very beginning of an epidemic, the infected individuals have not yet recovered meaning $N_{23} (1) = 0$. We deduce that~: $\hat{R}_{0P} = \hat{a}_P/\hat{c}_P = \hat{a}_P/0 = \infty$. This illustrates the lack of accuracy on the basic reproductive ratio during the initial phase of the outbreak.\vspace{1em}

\textbf{Remark 1~:} The unbiasedness property is specific to the case $T=1$. If

$T=2$, we have~:

$$
\hat{a}_P = \Frac{n [N_{12} (1) + N_{12} (2)]}{N_1 (0) N_2 (0) + N_1 (1) N_2 (1)}
$$

We deduce its expectation at date 1~:

$$
E_1 (\hat{a}_P) = n \Frac{N_{12} (1) + a N_1 (1) N_2 (1)}{N_1 (0) N_2 (0) + N_2 (1) N_2(1)},
$$

and by iterated expectation,

$$
E_0 (\hat{a}_P) = n E_0 \left[ \Frac{N_{12} (1) + a N_1 (1) N_2 (1)}{N_1 (0) N_2 (0) + N_1 (1) N_2 (1)}\right],
$$

\noindent which is the expectation of a complicated nonlinear function of counts $N_1 (1), N_2 (1), N_{12} (1)$.

\subsubsection{Binomial ML and Gaussian asymptotics}

This is the standard asymptotic theory when the Law of Large Numbers and  the Central Limit Theorem are applicable. The sample frequencies tend to their theoretical counterparts~: $\hat{p}_{jk} (t) \rightarrow p_{jk} (t)$, $\hat{p}_j (t) \rightarrow p_j(t), j,k= 1,2,3$, when $n$ tends to infinity. The ML estimators tend to the true parameter values~: $\hat{a} \rightarrow a, \hat{c} \rightarrow c, \hat{R}_0 = \hat{a}/\hat{c} \rightarrow a/c,$ at speed $1/\sqrt{n}$. $\hat{a}, \hat{c}$ are asymptotically independent, asymptotically normal and their variances are consistently estimated by~:

\begin{eqnarray}
  \hat{V} (\hat{a}) & = & \left\{ \Sum^T_{t=1} \left(\Frac{N_{11} (t) \hat{p}_2 (t-1)}{[1-\hat{a} \hat{p}_2 (t-1)]^2}\right) + \Frac{1}{\hat{a}^2} \Sum^T_{t=1} N_{12} (t)\right\}^{-1}, \\
  \hat{V}(\hat{c}) &=& \Frac{\hat{c} (1-\hat{c})}{\Sum^T_{t=1} N_2 (t-1)}.
\end{eqnarray}

\textbf{Remark 2~:} The Gaussian asymptotics can also be applied to other AML estimators as the Poisson AML. In this case the Poisson AML estimator of $a$ is still consistent, asymptotically normal. However, since the approximate log-likelihood is misspecified, its asymptotic variance is obtained by a sandwich formula, that involves the two expressions of the information matrix [see Huber (1967)].

\section{Monte-Carlo Study}
\setcounter{equation}{0}\def\theequation{4.\arabic{equation}}

Even when the Gaussian asymptotics can be used, we do not know if they are accurate for determining the confidence intervals on the different parameters $a,c,R_0$. In this section, we perform a Monte-Carlo analysis for some of the estimators introduced in Section 3. We fix the design as follows~:

$$ N_1 (0) = 3 000 000, N_2 (0) =  100, 1 000,T=20,c = 0.07, R_0 =2 $$

This corresponds to estimators computed on the period $[0,T]$. Note that the process of marginal counts is Markov. Therefore it also applies to a rolling estimator computed on $(t,t+T)$, say, where the marginal counts at $t$ are the counts fixed for $N_1 (0), N_2 (0)$. This explains why we allow for a large value of $N_2 (0)$ in the design. Figures 4 and 5 correspond to the parameters estimated by Approximated Poisson likelihood with $N_2 (0) = 100, 1000,$ respectively. They provide the finite sample distributions of parameters $a, c, R_0 = a/c$.\vspace{1em}

\newpage
Figure 4: Distributions of Approximate Poisson Estimators $ N_2 (0) = 100$.\vspace{1em}

\makebox[\textwidth]{
\begin{tabular}{cc}
  \includegraphics[width=80mm]{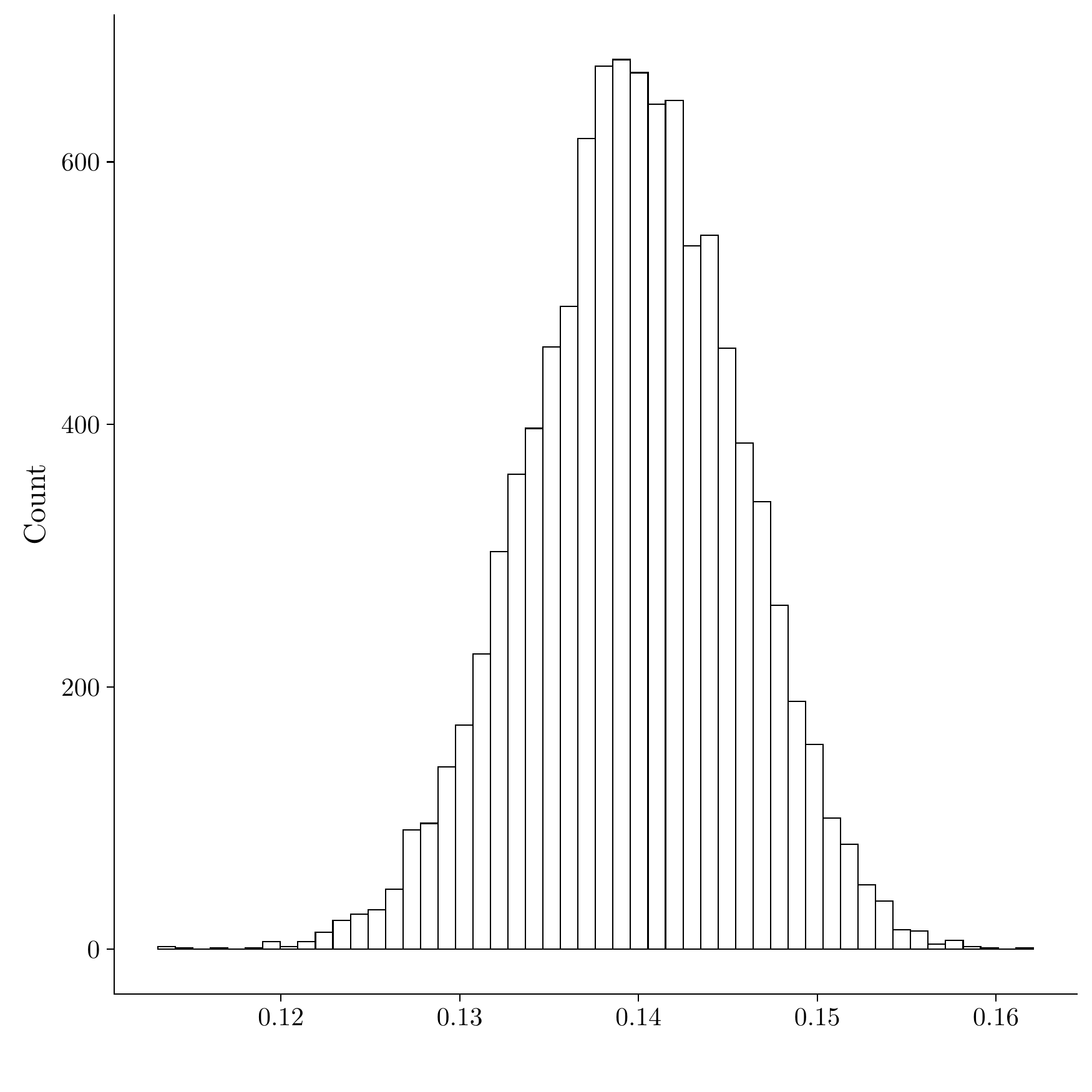} &   \includegraphics[width=80mm]{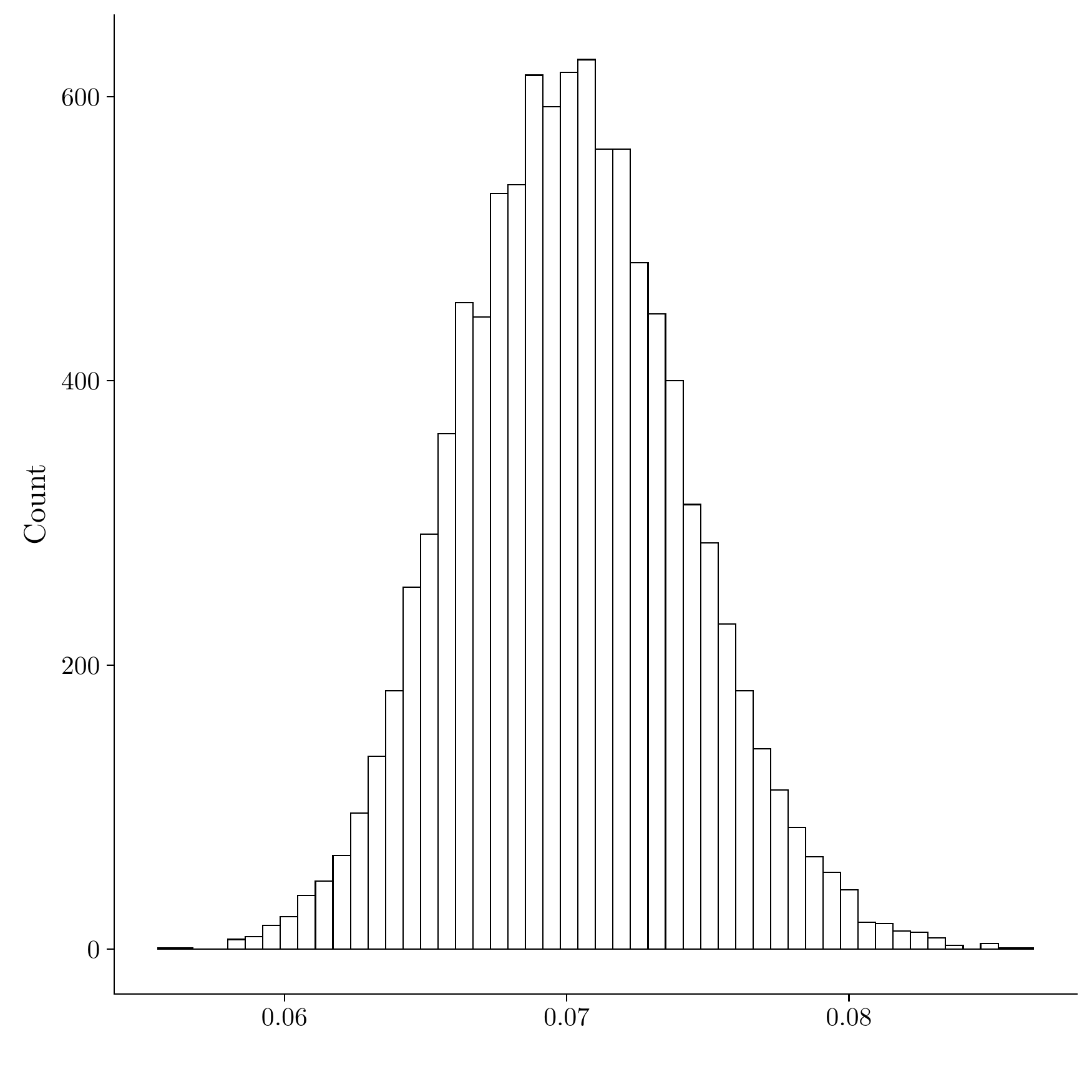} \\
$\hat{a}$ & $\hat{c}$ \\[6pt]
\multicolumn{2}{c}{\includegraphics[width=80mm]{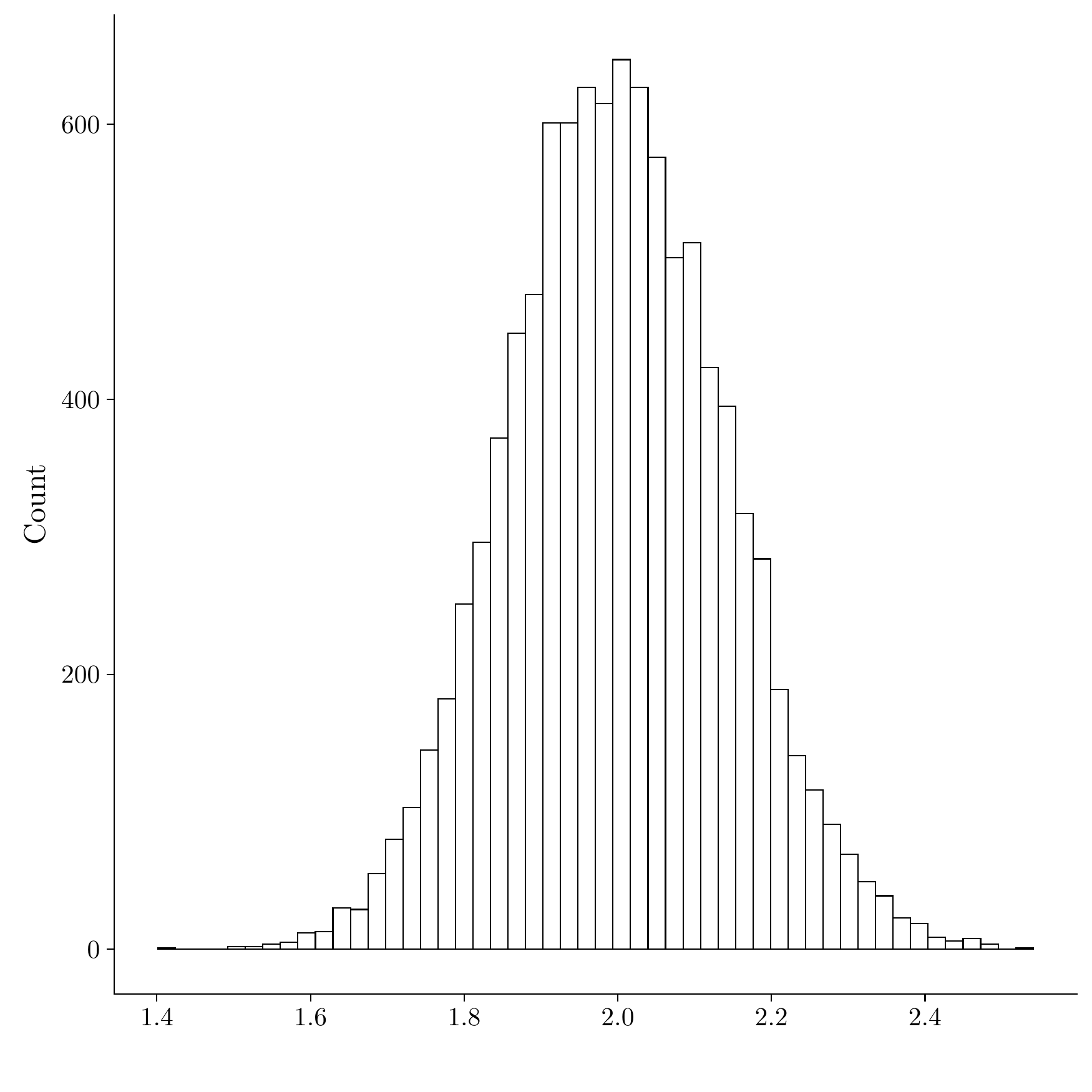} }\\
\multicolumn{2}{c}{$\hat{R}_0$}
\end{tabular}}

\newpage
Figure 5: Distributions of Approximate Poisson Estimators $ N_2 (0) = 1000$.\vspace{1em}

\makebox[\textwidth]{
\begin{tabular}{cc}
  \includegraphics[width=80mm]{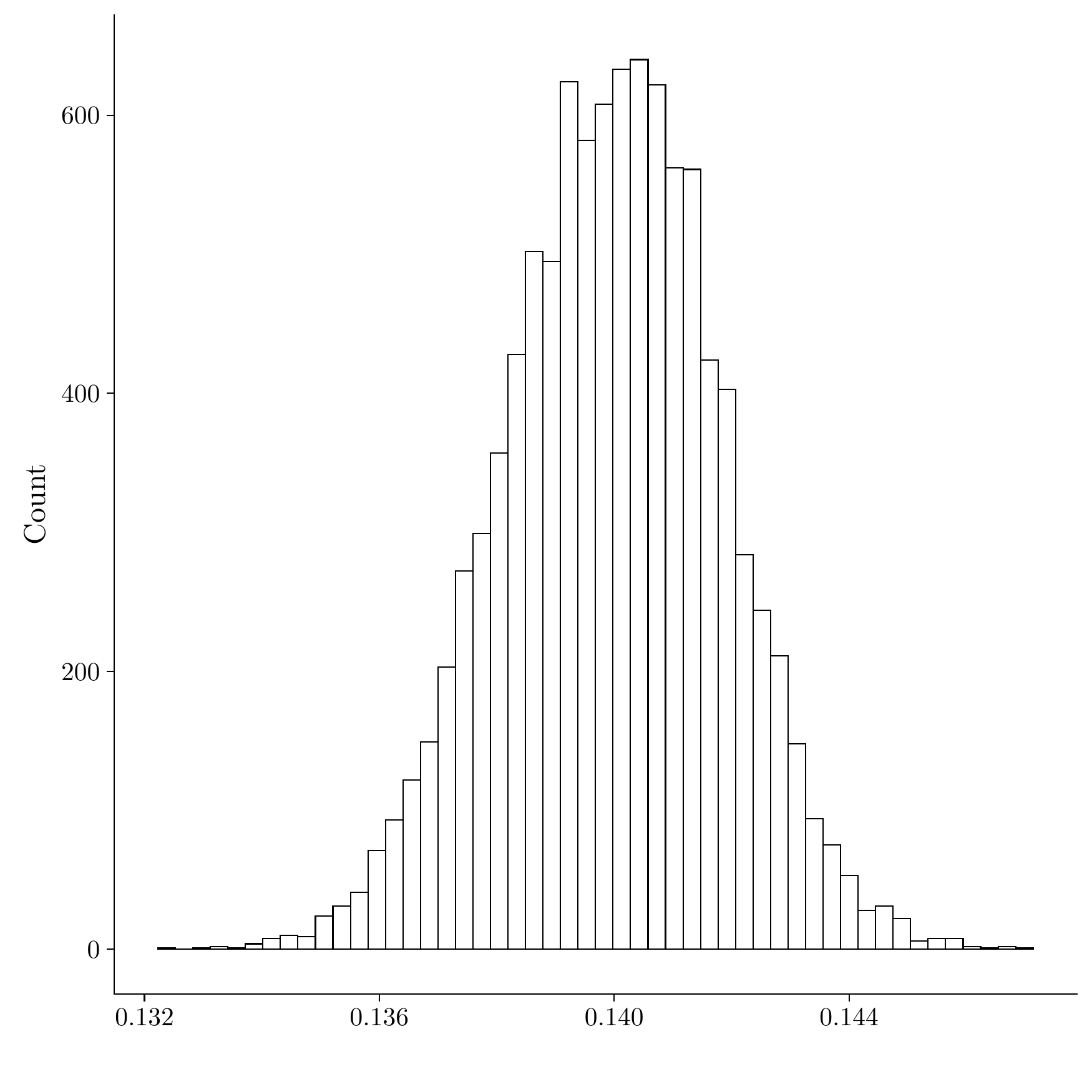} &   \includegraphics[width=80mm]{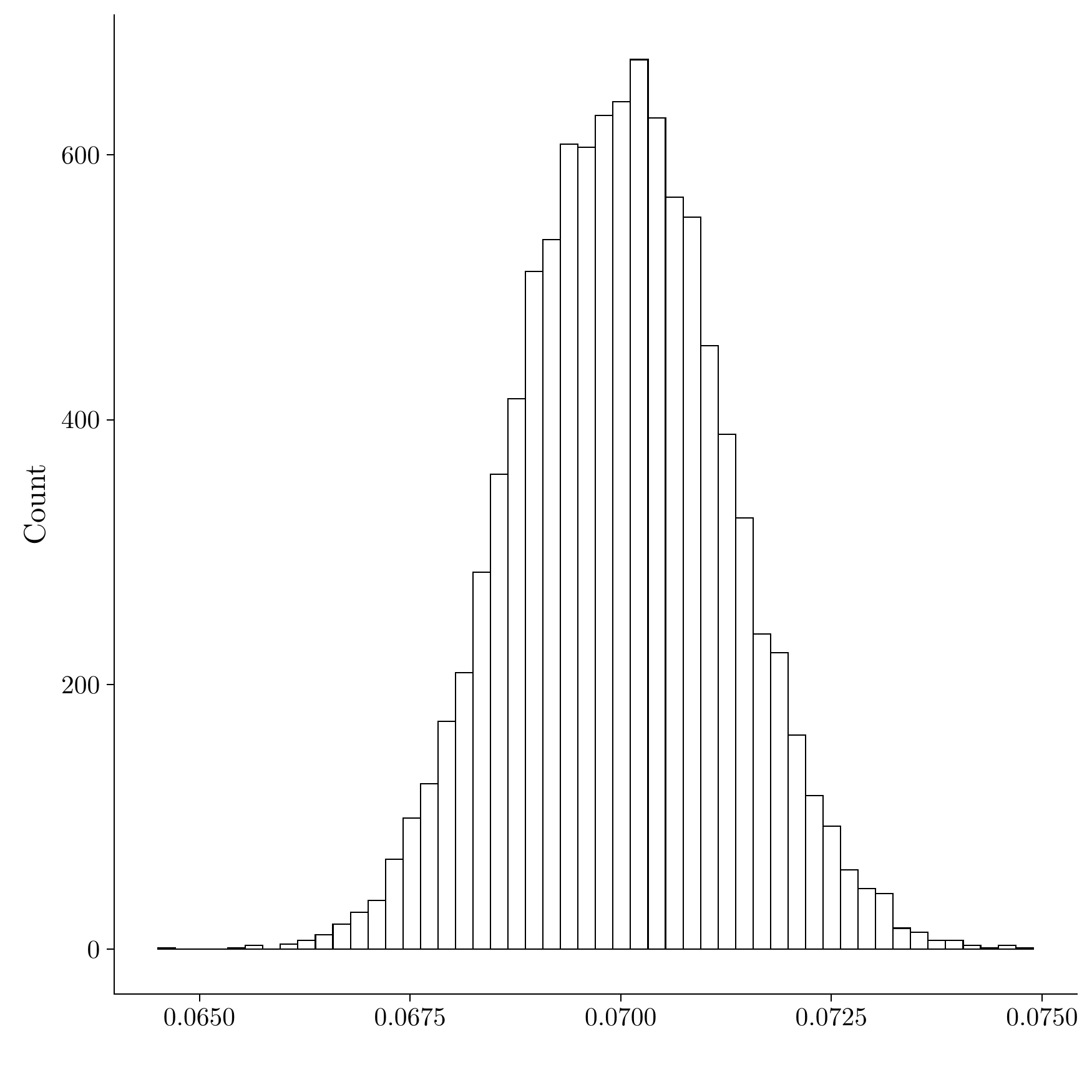} \\
$\hat{a}$ & $\hat{c}$ \\[6pt]
\multicolumn{2}{c}{\includegraphics[width=80mm]{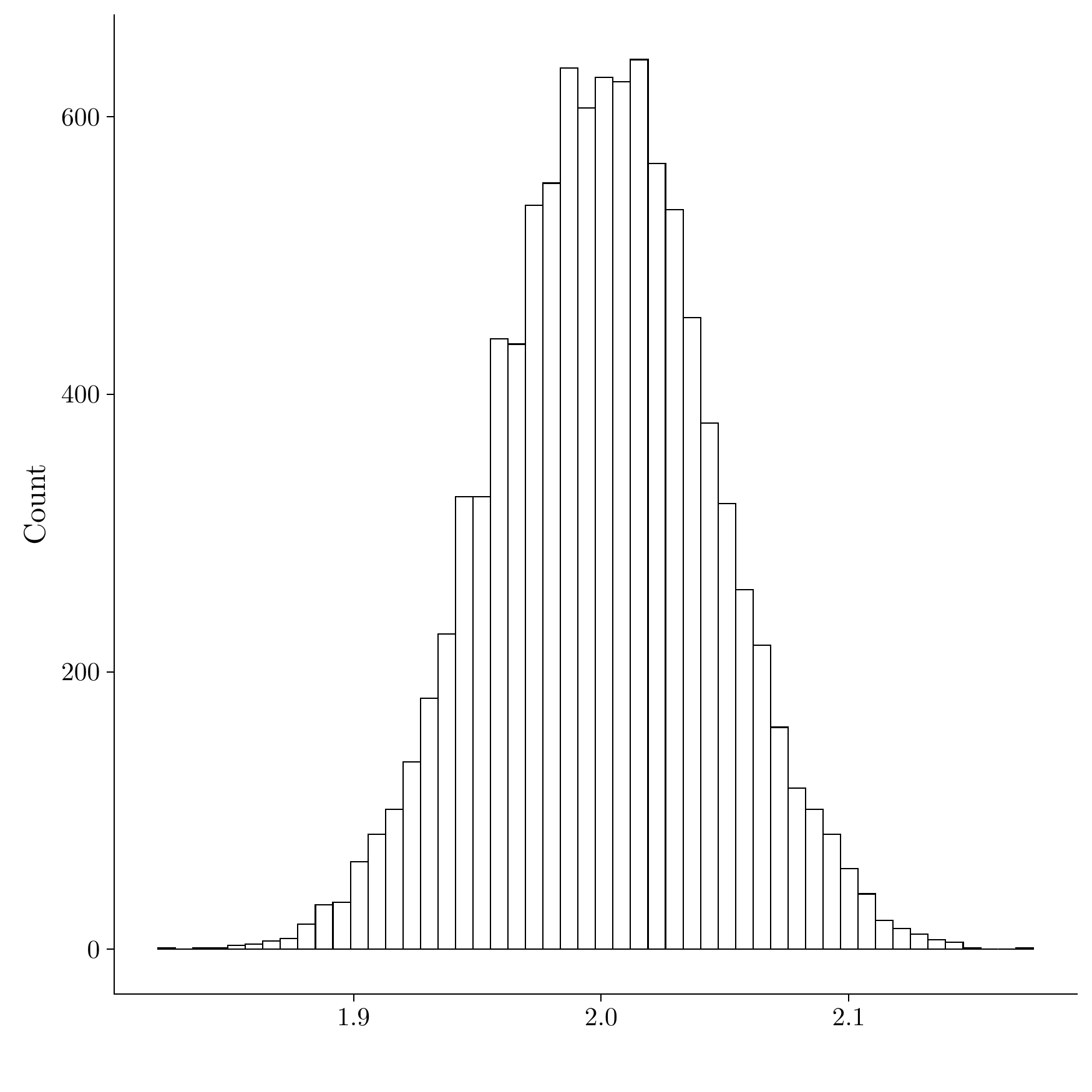} }\\
\multicolumn{2}{c}{$\hat{R}_0$}
\end{tabular}}

\newpage
\begin{tabular}{rrrrrrrrr}
  \hline
$N_2(0)$ & $T$ & $a$ & $c$ & $R_0$ & $\text{mean}(\hat{a})$ & $\text{var}(\hat{a})$& $\text{median}(\hat{a})$ & $\rho(\hat{a},\hat{c})$ \\
  \hline
5 & 20 & 0.035 & 0.07 & 0.5 & 0.03115 & 0.00045922 & 0.03044 & -0.112 \\
  5 & 20 & 0.140 & 0.07 & 2.0 & 0.13119 & 0.00099481 & 0.13539 & -0.246 \\
  5 & 40 & 0.105 & 0.07 & 1.5 & 0.09677 & 0.00051789 & 0.10100 & -0.380 \\
  5 & 40 & 0.140 & 0.07 & 2.0 & 0.13326 & 0.00044708 & 0.13732 & -0.489 \\
  100 & 20 & 0.140 & 0.07 & 2.0 & 0.13969 & 0.00003447 & 0.13973 & -0.005 \\
  100 & 40 & 0.070 & 0.07 & 1.0 & 0.06963 & 0.00001785 & 0.06977 & 0.006 \\
  200 & 20 & 0.070 & 0.07 & 1.0 & 0.06982 & 0.00001722 & 0.06986 & -0.027 \\
  200 & 40 & 0.070 & 0.07 & 1.0 & 0.06982 & 0.00000899 & 0.06990 & -0.009 \\
  300 & 20 & 0.070 & 0.07 & 1.0 & 0.06994 & 0.00001169 & 0.07000 & -0.008 \\
  300 & 40 & 0.035 & 0.07 & 0.5 & 0.03492 & 0.00000545 & 0.03496 & 0.000 \\
   \hline
\end{tabular}
\vspace{1em}

\centerline{Table 3 : Random Selection of $\hat{a}$ Summary Statistics}\vspace{1em}

\begin{tabular}{rrrrrrrrr}
  \hline
$N_2(0)$ & $T$ & $a$ & $c$ & $R_0$ & $\text{mean}(\hat{c})$ & $\text{var}(\hat{c})$& $\text{median}(\hat{c})$ & $\rho(\hat{a},\hat{c})$ \\
  \hline
50 & 40 & 0.035 & 0.07 & 0.5 & 0.07091 & 0.00006506 & 0.07034 & -0.004 \\
  100 & 40 & 0.070 & 0.07 & 1.0 & 0.07034 & 0.00001723 & 0.07015 & 0.006 \\
  100 & 40 & 0.105 & 0.07 & 1.5 & 0.07019 & 0.00000806 & 0.07008 & -0.007 \\
  200 & 20 & 0.105 & 0.07 & 1.5 & 0.07010 & 0.00001175 & 0.07000 & -0.007 \\
  200 & 20 & 0.140 & 0.07 & 2.0 & 0.07007 & 0.00000809 & 0.07004 & -0.007 \\
  300 & 20 & 0.035 & 0.07 & 0.5 & 0.07012 & 0.00001469 & 0.07008 & -0.004 \\
  500 & 20 & 0.035 & 0.07 & 0.5 & 0.07005 & 0.00000902 & 0.07002 & -0.003 \\
  500 & 20 & 0.105 & 0.07 & 1.5 & 0.07005 & 0.00000461 & 0.07000 & 0.008 \\
  500 & 40 & 0.035 & 0.07 & 0.5 & 0.07010 & 0.00000609 & 0.07002 & 0.012 \\
  1000 & 20 & 0.035 & 0.07 & 0.5 & 0.07006 & 0.00000433 & 0.07004 & 0.006 \\
   \hline
\end{tabular}
\vspace{1em}

\centerline{Table 4 : Random Selection of $\hat{c}$ Summary Statistics}\vspace{1em}

\begin{tabular}{rrrrrrrr}
  \hline
$N_2(0)$ & $T$ & $a$ & $c$ & $R_0$ & $\text{mean}(\hat{R}_0)$ & $\text{var}(\hat{R}_0)$& $\text{median}(\hat{R}_0)$ \\
  \hline
5 & 40 & 0.035 & 0.07 & 0.5 & 0.43255 & 0.08842763 & 0.42888 \\
  50 & 20 & 0.035 & 0.07 & 0.5 & 0.49951 & 0.01497913 & 0.49202 \\
  50 & 20 & 0.140 & 0.07 & 2.0 & 1.99277 & 0.04051883 & 1.98941 \\
  100 & 20 & 0.070 & 0.07 & 1.0 & 0.99856 & 0.01403924 & 0.99422 \\
  100 & 40 & 0.070 & 0.07 & 1.0 & 0.99327 & 0.00689571 & 0.99369 \\
  200 & 20 & 0.105 & 0.07 & 1.5 & 1.49857 & 0.00917040 & 1.49868 \\
  300 & 40 & 0.035 & 0.07 & 0.5 & 0.49858 & 0.00160204 & 0.49925 \\
  500 & 40 & 0.035 & 0.07 & 0.5 & 0.49956 & 0.00096347 & 0.49970 \\
  500 & 40 & 0.070 & 0.07 & 1.0 & 0.99871 & 0.00137583 & 0.99869 \\
  1000 & 20 & 0.070 & 0.07 & 1.0 & 0.99950 & 0.00137986 & 0.99882 \\
   \hline
\end{tabular}

\centerline{Table 5 : Random Selection of $\hat{R}_0$ Summary Statistics}
\newpage

Whereas a significant skewness is observed for the estimation of contagion parameter, this feature largely disappears for the reproduction number. This is due to the nonlinear transformation to compute $R_0$, but also to the dependence between $\hat{a}$ and $\hat{c}$. $R_0$ is known at $\pm 20\%$ for $N_2 (0) = 100,$ at $\pm 10\%$ for $N_2 (0) = 1000.$

To have more insight on the finite sample properties of these estimators, we provide summary statistics including the correlation ($\rho$) between $\hat{a}$ and $\hat{c}$, for different designs (a,c), initial $N_2 (0)$, and number of observations $T$ in Tables 3-5. Finite sample distributions for the estimators computed by unfeasible Gaussian approximate likelihood are given in Appendix 3.

\section{The Reproductive Number Under Heterogeneity}

\subsection{Model with heterogeneity}

Another source of variability for the estimated $R_0$ is due to latent heterogeneity and concerns the definition of $R_0$ itself. For illustration, we consider a situation with two homogeneous populations, population 1 and population 2, say. Then the SIR model to a (SIR)$^2$ model with six states~: $S_1 I_1 R_1 S_2 I_2 R_2$ in the terminology of Gourieroux, Jasiak (2020)b, Appendix 1. The (6,6) transition matrix is block diagonal with diagonal blocks given by~:

$$
P_{j,t} = \left( \begin{array}{ccl}
1-a_{j1} \Frac{N^1_2 (t-1)}{N^1} - a_{j2} \Frac{N^2_2 (t-1)}{N^2} & a_{j1} \Frac{N^1_2 (t-1)}{N^1} + a_{j2} \Frac{N^2_2 (t-1)}{N^2} & 0 \\ \\
0 & 1-c_j & c_j \\ \\
0 & 0 & 1
\end{array}
\right),
$$

\noindent for $j=1,2,$ where $N^j_2 (t)$ (resp. $n^j$) is the number of infected people in population $j$ (resp. the size of population $j$). Typically, the two populations can correspond to two age categories, young and old, say. Now the contagion parameter has a matrix form~: $A = \left( \begin{array}{cc}a_{11} & a_{12} \\ \\ a_{21} & a_{22} \end{array}\right).$ Indeed there is contagion within each population~: $a_{11}, a_{22}$, and between the populations $a_{12}, a_{21}$.

The (SIR)$^2$ model can be constrained by introducing degrees of infectiveness and of infection vulnerability, denoted $\alpha_j$ and $\beta_j$, respectively. Then, the contagion matrix $A$ is equal to~: $A=\beta \alpha'$. This matrix has reduced rank equal to 1.\vspace{1em}

The existence of between and within contagions will modify the notion of the reproductive number which now must account for the different types of contagions from a new infected individual of type 1 (resp. 2) to individuals at risk of either type 1, or 2. The initial reproductive number now has a matrix form~:

$$
\begin{array}{lcl}
               R_{0,0} & = &\beta \tilde{\alpha}', \\ \\
 \mbox{with} \; \tilde{\alpha}_j &  = & \alpha_j/c_j, j=1,2.
\end{array}
$$

The diagonal elements of matrix $R_{0,0}$ can be very different. For instance, if one segment includes the super-spreaders, the reproductive number can pass from a value around 2 [WHO (2020)] to a value between 4.5 and 11.5 [Kochanczik et al. (2020)].

\subsection{Omitted heterogeneity}

Let us now assume such an underlying (SIR)$^{2}$ model and aggregate the two subpopulations in $S=S_1 U S_2$, $I=I_1 U I_2, R=R_1 U R_2$. There is an aggregation bias that implies that the cross-sectional counts~:

$$
N_1(t) = N^1_1 (t) + N^2_1 (t), N_2 (t) = N^1_2 (t) + N^2_2 (t), N_3 (t) = N^1_3 (t) + N^2_3 (t),
$$

\noindent no longer define a Markov process. However, it is still possible to compute the transition matrix at horizon 1. Let us for instance consider the probability for an individual at risk at date $t-1$ (i.e. in state $S$ at $t-1$) to be infected at date $t$ by a new infectious individual. By the Bayes formula, we get :\vspace{1em}

 $P$[infected at $t$ $|$ at-risk at $t-1$]

$= P[$ infected at $t$ $|$  at-risk $t-1$, in Pop 1] $P$ [ at-risk $t-1$, Pop 1 $|$ at-risk at $t-1$]

+ $P$ [infected at $t$ $|$ at risk at $t-1$, in Pop 1] P[at risk at $t-1$, in Pop 2 $|$ at risk at $t-1$]

$$
\begin{array}{ll}
  = & \Frac{N^1_1 (t-1)}{N_1 (t-1)} \left[ a_{11} \Frac{N^1_2 (t-1)}{N^1} + a_{12} \Frac{N^2_2 (t-1)}{N^2}\right] \\ \\
  + & \Frac{N^2_1 (t-1)}{N_1 (t-1)} \left[ a_{21} \Frac{N^1_2 (t-1)}{N^1} + a_{22} \Frac{N^2_2 (t-1)}{N^2}\right] \\ \\
  = & \left[ \beta_1 \Frac{N^1_1 (t-1)}{N_1 (t-1)} + \beta_2 \Frac{N^2_1 (t-1)}{N_1 (t-1)}\right] \left[ \alpha_1 \Frac{N^1_1 (t-1)}{N^1} + \alpha_2 \Frac{N^2_2 (t-1)}{N^2}\right] \\ \\
  = & a_t \Frac{N_2 (t-1)}{N},
\end{array}
$$

\noindent where $a_t$ is the dated transmission parameter in the SIR model with omitted heterogeneity. Therefore, using the standard SIR model when there is heterogeneity implies a time varying contagion parameter. A similar effect, known as the mover-stayer phenomenon, exists for the intensity to recover from the infection state and leads to a time varying $c_t$, and therefore on the reproductive number : $R_{0,0,t} = a_t/c_t$.\vspace{1em}

This type of decomposition can easily be extended to more  than two homogeneous subpopulations [see e.g. Alipoor, Boldea (2020)].

\section{Instantaneous Reproductive Number}
\setcounter{equation}{0}\def\theequation{6.\arabic{equation}}

There exists on the market different packages to estimate a reproductive number, usually in a rolling way. We discuss below one set of estimation methods to approximate the instantaneous reproductive number, a notion that differs from the basic reproduction number.\footnote{A proposed alternative is to define $R$ as an exponential rate of diffusion of the disease usually estimated by either log-regression, or Poisson regression. [see e.g. Lipsitch et al (2003), Wallinga, Lipsitch (2007), Boelle et al. (2009)]. Other approaches are based on some assumption of a network of contagion, as in Wallinga, Teunis (2004). However their proposed methodology assumes a static equilibrium network, tries to reconstitute a tracing ex-post, without really taking into account the dynamic of the disease. This implies a right censoring bias [Cauchemez et al. (2006)]} This type of computation and the associated software can be found in [Cori et al. (2013), with the EpiEstim package, the time dependent reproduction number in the RO package Obadia et al. (2012)] and are used, for instance, in the official reproductive number provided by Public Health Ontario [PHO (2020)]. Even if this approach is presented to estimate time varying reproduction number, the methodology is expected to work also in a framework of a weakly time dependent reproduction number. This is why the discussion is done under the SIR model.

\subsection{A generic estimator}

Alternative estimation approaches of the reproductive number have been introduced in the literature and in the software. They are often presented as almost model free and are popular among practitioners since they are simple to use. An example of such a generic approach has been introduced in Fraser (2007), Cori et al. (2013) following a similar idea appeared in Wallinga, Teunis (2004), p511. The method requires the knowledge of the sequence of new infections only, $N_{12} (t)$, with $t$ varying. The count of time $t$ is written on the lagged counts as~:

$$
N_{12} (t) \simeq \Sum^S_{s=1} \gamma_s N_{12} (t-s),
$$

\noindent and the regression coefficients can be normalized as~: $\gamma_s = w_s \gamma$, where $\Sum^S_{s=1} w_s = 1.$\vspace{1em}

The estimated ``instantaneous reproduction number" is defined in EpiEstim as [see Cori et al. (2020), p2]~:

\begin{equation}
  \hat{R}^i_t = \Frac{N_{12} (t)}{\Sum^t_{s=1} N_{12} (t-1) \hat{w}_s},
\end{equation}

\noindent where the sum in the denominator starts at  the first time of infections and $\hat{w}_s$ is a Bayesian estimate of the infectiousness profile. \footnote{Sometimes the infectiveness profile of $w_s$ is even not really estimated, but fixed ex-ante, possibly through a prior [see e.g. Cory et al. (2013), webappendix 4 and the discussion below]. The results will significantly depend on this selected sequence.}

Such a simple procedure is not necessarily robust: it depends on the length of the estimation period, the number of lags in the sum appearing in the denominator, on the choice of the infectiveness profile $w_s$ and on its estimate. But more importantly, any generic approach will work well under some implicit assumptions and if the notion on interest is correctly defined under these assumptions.

Let us illustrate the properties of the EpiEstim approach. This estimator is usually computed in a rolling way. It is based on a Bayesian assumption with a prior on the distribution of the serial interval, that is, the time from symptom onset in a primary case (infector) to symptom onset in a secondary case (infectee). This prior depends on two parameters, that are  a mean and a standard deviation. In EpiEstim1 we have retained the same log-normal prior : mean = 4.5 days, standard deviation = 2.5 days chosen by PHO (2020). It is close to the prior in Nishura et al (2020) [mean = 4.7 days, standard deviation = 2.5, based on 18 pairs of infector - infectee], but different from the prior in Du et al. (2020) [mean = 3.96, standard deviation = 4.15, based on 468 pairs].

In Figure 6 we display different estimates computed from a simulated series satisfying the SIR model. The EpiEstim1 estimate is calculated on a window of seven days. The approximate ML estimates [Binomial, Poisson, Unfeasible Gaussian] are computed at each date $t$ using all the data from the outbreak. The Poisson and Binomial estimates cannot be distinguished. All estimates have poor properties at the beginning, when the number of new infections is rather small and there are almost no recoveries. The ML estimators show a variability which becomes rather small after 30 days, and they converge to the true value of the basic reproductive number.

Let us now discuss the evolution of the EpiEstim1 estimator. This evolution is strongly dependent on the Bayesian approach that is used. Even if the estimate is computed in a rolling way, only seven observations are taken into account at each date $t$, that gives a significant weight to the prior. This explains the weak variability of this estimate over time. Moreover the level of the estimate is strongly dependent on the selected prior and clearly it is not varying around the true value of $R_0$, even if it accounts for the information in the counts of new infected. In EpiEstim1, we  have followed the current practice in which the prior relies on preexisting estimates of the serial interval distribution. These estimates can correspond to another disease, to the same disease in another country, or in the case of COVID-19, a small number of observations: 18 pairs in Nishura et al. (2020), endogenously selected (12 among these pairs correspond to transmission within family, then to short transmissions). These are estimated using the definition of the serial interval as the time between symptomatic cases [Thompson et al. (2019)] which will underestimate the mean and uncertainty due to the presence of asymptomatic infection periods and/or individuals. 

Finally the choice of a log-normal prior instead of a gamma prior, that is, of a thin tail prior instead of a fat tail prior can also lead to an underestimation of the level. A further implication of the Bayesian approach can be observed when one provides the software with a zero vector instead of a vector containing new infections. In this scenario, the process will simply return a reproduction number which is constant over time.

 In order to check the role of the prior, we also display in Figure 6 the plot corresponding to the EpiEstim estimator with a log-normal prior with the same mean and standard deviation as the geometric distribution with mean 14 days. This is an unfeasible estimator assuming that the infectivity profile is fixed at its true value [see the discussion in 6.4.2, and formula (6.19)]. A convergence to the true value $R_0$ is now observed. These drawbacks of the EpiEstim approach have been recently mentioned by some authors of the R software package [Thompson et al. (2019)], who propose an improved version of their package. We will discuss it later on, however this recent version has not yet been implemented. \\

\newpage
\textbf{Figure 6 : Comparison Using EpiEstim on Simulated SIR Model Data}

\begin{flushleft}
\includegraphics[width=\linewidth]{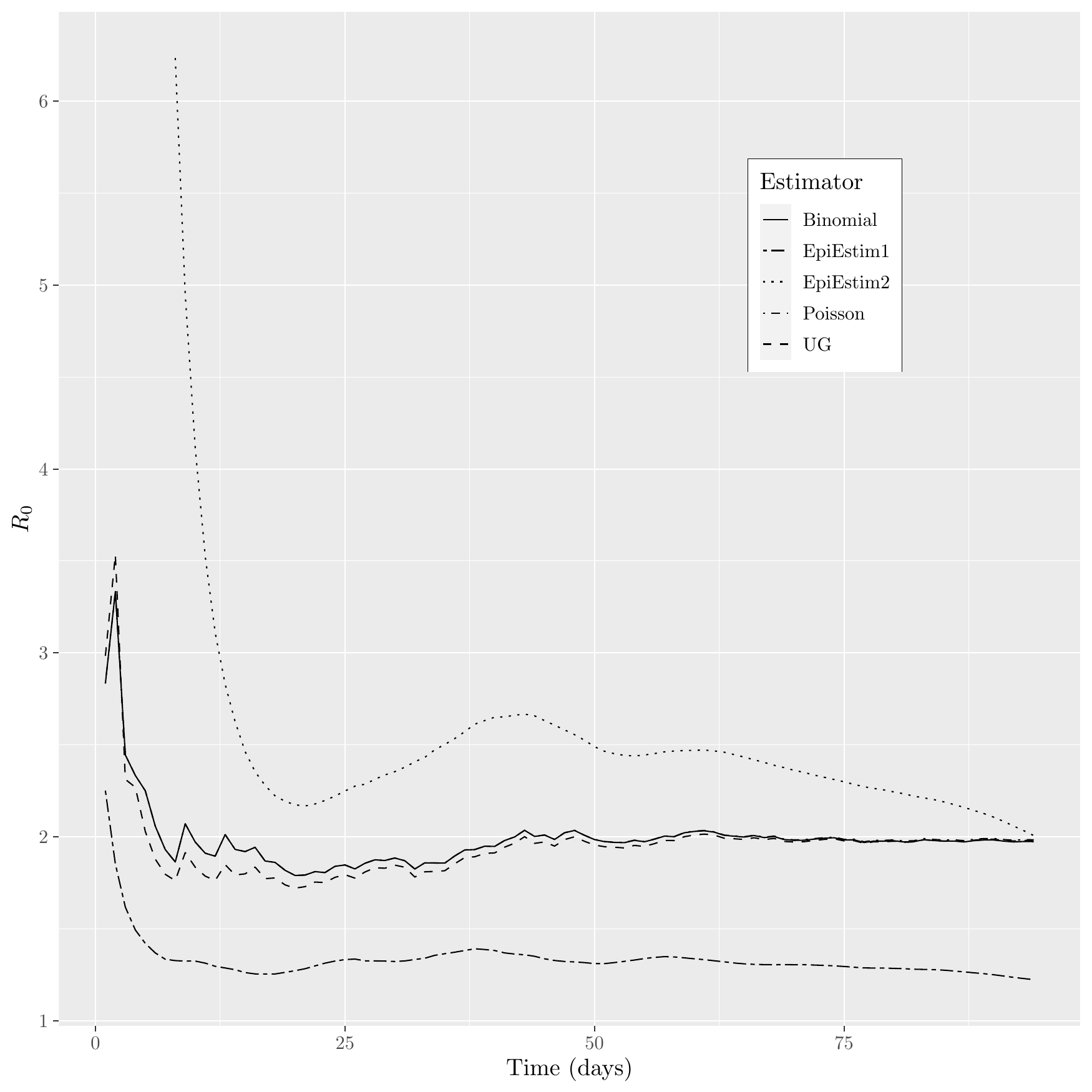}
\end{flushleft}

The objective of the following sections is to discuss the origin of the EpiEstim approach in order to explain the differences in the estimates observed in Figure 6.

\subsection{The underlying model}

To understand formula (6.1), we have to extend the basic SIR model. We retain a constant contagion parameter $a$ but introduce a stochastic duration of infectiousness D, which is not necessarily geometrically distributed. Its distribution is characterized by its survival function denoted:

\begin{equation}
  \gamma (s) = P[D \geq s], s=1,2,...
\end{equation}

Then the expression of the basic reproductive number is easily derived (see Section 2.3). It becomes~:

\begin{equation}
R_{0,t} = \Frac{a}{N_1 (t)} \Sum^\infty_{s=0} \{ E_t (N_1 (t+s) \gamma (s)\}.
\end{equation}

Let us now write this expression in terms of new infections. We have~:

\begin{equation}
  N_1(t) - N_1 (t-1) = -N_{12} (t),
\end{equation}

\noindent and then~:

\begin{equation}
N_1 (t+s) = N_1 (t) - \Sum^s_{k=1} N_{12} (t+k).
\end{equation}

By replacing $N_1 (t+s)$ by this expression in equation (6.3), we get~:

$$
\begin{array}{lcl}
R_{0,t} & = & \Frac{a}{N_1 (t)} \Sum^\infty_{s=0} \{ \gamma(s) [N_1(t) - E_t [\Sum^s_{k=1} N_{12} (t+k)]\}\; \mbox{(with the convention}\; \Sum^0_{k=1} = 0)\\ \\
&=& a \Sum^\infty_{s=0} \gamma (s) - \Frac{a}{N_t (t)} \Sum^\infty_{s=1} \Sum^s_{k=1} [\gamma (s) E_t (N_{12} (t+k))] \\ \\
&=& a \Sum^\infty_{s=0} \gamma (s) - \Frac{a}{N_1 (t)} \Sum^\infty_{k=1} [E_t N_{12} (t+k) \Sum^\infty_{s=k} \gamma (s)].
\end{array}
$$

The partial sums of the survival function $\gamma (s)$ can be rewritten in terms of moments of the stochastic duration of infectiousness. We get~:

\begin{equation}
  R_{0,t} = a E (D) - \Frac{a}{N_1 (t)} \sum^\infty_{k=1} \{ E [(D-k)^+] E_t (N_{12} (t+k)]\},
\end{equation}

\noindent where $x^+ = Max (x,0).$\vspace{1em}

\textbf{Remark 3 :} In the standard SIR model, formula (6.6) becomes~:

$$
R_{0,t} = (a/c) \{ 1-\Sum^\infty_{k=1} [(1-c)^k E_t (N_{12} (t+k))]\}.
$$

Let us now discuss the conditional expectation $E_t$. In the SIR framework, the conditioning set includes the current and lagged values of the $N_{jk} (t), j,k=1,2,3,$ or equivalently of the cross-sectional counts $N_k (t), k=1,2,3.$ Therefore the sufficient summary of the past information requires two sequences of counts.

By considering a single sequence of counts, i.e.
the counts of new infected people, the generic approach is changing the information set and modifies the definition of the dated reproductive number (see the discussion in Section 6.3).

With this restricted information set,  the new reproductive number is~:

\begin{equation}
  R^N_{0,t} = a E D - \Frac{a}{N_1 (t)} \Sum^\infty_{k=1} \{ E [(D-k)^+] E [N_{12} (t+k) | \underline{N_{12} (t)}]\},
\end{equation}

\noindent where index $N$ indicates the restriction to new infections.

Can we expect a linear prediction formula for the prediction of the counts of new infected people, such as~:

\begin{equation}
  E [N_{12} (t+k)| \underline{N_{12} (t)}] = \Sum^\infty_{h=0} \beta_{kh} N_{12} (t-h),
\end{equation}

\noindent with time independent $\beta_{kh}$ coefficients? Likely not, due to the nonlinear dynamics of a contagion model and its analysis during a non-stationary episode.

\subsection{Which Definition of Reproduction Number}

To understand the significant difference between the formula (6.1) for $\hat{R}^i_t$ and the formula (6.7) for $R^N_{0t}$, it is useful to come back on the paper in which the notion of instantaneous reproductive number was introduced [see Fraser (2007)]. Fraser's approach is based on a renewal equation~:

\begin{equation}
  I(t) = \Sum^\infty_{s=1} \beta (t,s) I (t-s),
\end{equation}

\noindent where $I(t)$ is the incidence proportion \footnote{See CDC (2012) for the different definitions of incidence depending on the selected denominator.} at $t$, or attack rate (approximated by $N_{12} (t)/N_1 (t-1))$ and $\beta (t,s)$ is the effective contact rate between infectious and susceptible individuals taking into account the generation of new infected people. Both the SIR model and the renewal equation appear in the same paper of Kermack-McKendrick (1927) and are compatible. Under the SIR model, the contact rate $\beta (t,s)$ is a complicated nonlinear function of the sufficient summary counts, that is, the new infected and new recovered counts between dates $t-s$ and $t$. Therefore, in the SIR framework, the renewal equation (6.9) involves a ``lagged endogenous" contact rate, in fact an equilibrium contact rate.

Let us now give the definitions of reproduction ratios in Fraser (2007). Two notions called ``case reproductive ratio" and ``instantaneous reproductive ratio", respectively, are introduced with the main objective to get a ready-to-use measure based on simple analytical formulas. It is important to note that they have new names, since they significantly differ from the standard basic and effective reproductive numbers. It is particularly important to note that they do not have the same interpretation. For instance, the instantaneous reproductive number is defined from (6.9) by considering what reproduction can be expected if ``the conditions remain unchanged" in the past, i.e. $I(t-s) = I, s=1,2$. The ratio is then defined as [see eq.(3) in Fraser (2007)]~:

\begin{equation}
  R^i_t = \Frac{I_t}{I} = \Sum^\infty_{s=1} \beta (t,s).
\end{equation}

This practice disregards the endogeneity of the contact rates. Indeed the contact rates also depend on the evolution of the number of new infected individuals which has assumed to be unchanged in the ``linear" component of the renewal equation but not in the (nonlinear) contact rate. Moreover, the assumption of unchanged condition is not necessarily compatible with the evolution with peak corresponding to a SIR model and to the observations of $I(t)$, or $N_{12} (t)$. In fact, one objective of this definition was to reveal in the measure the expected sudden decrease of $R$ resulting from a new effective control undertaken at time $t$.

Finally to derive the expression (6.1), it is also assumed a decomposition of the contact rate as~:

\begin{equation}
  \beta (t,s) = R^i_t w(s),
\end{equation}

\noindent where the $w(s), s=1,\ldots,S$ sum up to $1$.

By taking into account this reduced rank condition, the renewal equation (6.9) is equivalent to~:

\begin{equation}
  R^i_t = I_t/\Sum^S_{s=1} (I_{t-s} w (s)],
\end{equation}

\noindent which explains the generic estimate (6.1) (if $N_1 (t)$ is not changing a lot, see the discussion in Section 6.4) and its interpretation as the ratio of new infections by the total infectiousness of infected individuals up to time $t-1$.

\subsection{Sources of Bias}

Let us now make explicit the three main sources of bias when a formula such as (6.12) is used to approximate the basic reproductive number. The discussion is done under the assumption \footnote{Some Monte-Carlo studies have been performed in the literature under some specific renewal model, comparing $R^i_t$ with its estimate [see e.g. Cori et al. (2013)]. Such an analysis is misleading since the right comparison is between the estimate of $R^i_t$ and the basic $R_{00}$ to measure the bias that can result from the approximations described in Section 6.3.} of a SIR model with constant parameters $a,c, R_{00}=a/c.$ As usual in epidemiology it is important to distinguish between the stochastic models of the observations and its associated deterministic (or mechanistic) model corresponding to a virtual population of infinite size [Breto et al. (2009), Smieszek (2009), Funk et al. (2018)].

\subsubsection{The mechanistic model}

Let us derive a mechanistic model of infection derived from the SIR model. As in Section 3.2, we denote $p_1 (t), p_{12} (t)$ the theoretical probabilities corresponding to the frequencies $\hat{p}_1 (t) = N_1 (t)/n, \hat{p}_{12} (t) = N_{12} (t)/n.$ We assume that the frequencies $\hat{p}$ tends to the $p's$ when $n$ tends to infinity. In this case, $p(t)$ is also equal to the (unconditional) expectation of $\hat{p}(t)$. Let us focus on the mechanistic component of the model for infection, that is without considering recovery.

When $n$ varies, we need to appropriately adjust the contagion parameter to derive the mechanistic model, i.e. to replace $a$ by $a_n = a/n$, say. Then we have~:

\begin{equation}
  E_{t-1} \left( \Frac{N_{12} (t)}{n}\right) = a \Frac{N_{1} (t-1)}{n} \Frac{N_{2} (t-1)}{n}.
\end{equation}

Let us now decompose the count $N_2 (t-1)$ as~:

\begin{equation}
  N_2 (t-1)= \Sum^t_{s=1} N_2 (t-1;s),
\end{equation}

\noindent where $N_2 (t-1;s)$ is the number of individuals infected at $t-s$ for the first time and still infectious at $t-1$. In the SIR model with geometric duration of infection, we have~:

\begin{eqnarray}
   E_{t-1} \left[ \Frac{N_2 (t-1;s)}{n}\right] & =&\Frac{N_{12} (t-s)}{n} (1-c)^{s-1}, \\
   \mbox{then:}\;\; E \left[ \Frac{N_2 (t-1,s)}{n}\right] & = & (1-c)^{s-1} E \left(\Frac{N_{12} (t-s)}{n} \right).
\end{eqnarray}

Making $n$ tend to infinity in these relations and using the fact that the limit of the $p's$ are deterministic, we get the deterministic recursive equation~:

\begin{equation}
  p^*_{12} (t) = a p_1 (t-1) \Sum^t_{s=1} [(1-c)^{s-1} p^*_{12} (t-s)],
\end{equation}

\noindent or equivalently,

\begin{equation}
  p^*_{12} (t) = a [1-\Sum^t_{s=1} p^*_{12}  (t-s)] \Sum^t_{s=1} [(1-c)^{s-1} p^*_{12}  (t-s)],
\end{equation}

\noindent where $p^*_{12} (t) = \lim_{n\rightarrow \infty} [N_{12}(t)/n]$. $p^*_{12} (t)$ differs from $p_{12} (t),$ by the denominator $n$ instead of $N_1 (t-1)$, except at the beginning of the disease. From (6.18), we see that the series $p^*_{12}  (t) = E (N_{12} (t)/n)$ satisfies a quadratic recursive equation with an order that tends to infinity with $t$.

\subsubsection{The linearization bias}

A first approximation assumes that $p_1 (t-1)$ is close to 1. This approximation is reasonable and standard at the beginning of the disease, but can induce biases in the medium run (when looking for the peak) and in the long run (when looking for final size and herd immunity). Under this approximation, we get~:

\begin{eqnarray}
  p^*_{12}  (t) & \simeq & a \Sum^t_{s=1} [(1-c)^{s-1} p^*_{12}  (t-s)] \nonumber \\
  &=&  \Frac{a}{c} \Sum^t_{s=1} [w (s) p^*_{12}  (t-s)], \; \mbox{or}\nonumber \\
  p^*_{12}  (t) & = & R_{0,0} \Sum^t_{s=1} [w(s) p^*_{12}  (t-s)],
\end{eqnarray}

\noindent with $w(s) = c(1-c)^{s-1}.$\\

The relation (6.19) on the expected new infection rates is the basis of the methodology introduced in Fraser (2007).

\subsubsection{The causality bias}

In a deterministic equation as (6.19), the fact that a variable is in the right hand side, or in the left-hand side of the equation is not important. However this becomes important, when the population size is large, but finite, and the probabilities replaced by their frequency analogues. We can always deduce from (6.19) a relation on observations as~:

\begin{equation}
N_{12} (t) = - R_{0,0} \Sum^t_{s=1} [w(s) N_{12} (t-s)] + u(t),
\end{equation}

\noindent where $u(t)$ are errors, but there is no reason for the $u(t)'s$ to be independent of one another and more importantly, to be independent of the lagged $N_{12} (t-s), (s=1, \ldots)$. In fact, under the SIR model, they are dependent and correlated with the lagged counts. This can induce a bias when a (Poisson) least squares is applied to estimate $R_{0,0}$ (assuming either fixed $w(s)$, i.e. given $c$, or $w(s)$ estimated by OLS).

\subsubsection{The bias and lack of efficiency of the two-step approach}

Presenting an estimation approach of a simple OLS type and based only on the counts of new detected individuals has made some success for the EpiEstim approach. However, this presentation is a bit misleading. Indeed the weights $w(s)$ (i.e. parameter $c)$ are unknown. Either they are fixed in some arbitrary way (see the description in the EpiEstim package) and this bias on the weights will imply a bias in the computation of $R_0,$ as seen in the example of Figure 6, or  they have to be estimated. This estimation will require more complicated approaches and the efficient use of other count series such as the total count of infected individuals, or the counts of new recovered individuals (see Section 2.2), or alternatively, tracing data on infector/infectee pairs (see the discussion in the conclusion).

In fact, the mechanistic equation (6.19) shows that there is an identification issue for large population sizes and automatically a poor accuracy for an estimate based on the observation of counts $N_{12} (t)$ only. Indeed, by introducing the lag-operator $L$, equation (6.19) can be written as~:

$$
\begin{array}{llcl}
 & p^*_{12} (t) & = & R_{0,0} \Sum^t_{s=1} [c(1-c)^{s-1} L^s  p^*_{12} (t)] \\ \\
 \Leftrightarrow & p^*_{12} (t) & = & R_{0,0} c L (\Sum^{t-1}_{s=0} (1-c)^s L^s)  p^*_{12} (t) \\ \\
 &&\sim & R_{0,0} c L \Frac{1}{1-(1-c) L}  p^*_{12} (t) \;  \;\; \mbox{(for large}\; t)
\end{array}
$$

This is equivalent to~:

$$
\begin{array}{ll}
   & [1- (1-c) L]  p^*_{12} (t) = c R_{0,0} L  p^*_{12} (t) \\ \\
   \Leftrightarrow &  p^*_{12} (t) = [1+ c [R_{0,0} -1]  p^*_{12} (t).
\end{array}
$$

Due to the linearization highlighted in Section (6.4.2), the mechanistic model tends to an exponential pattern for large $t$. Moreover, for a large number of observation $T$, we can essentially identify the ``rate of explosion" equal to $1+c [R_{0,0} -1]$, not separately $R_{0,0}$ and $c$.

\subsection{The Autoregression Estimate}

An alternative estimate of the reproductive number can also be introduced based on the approximate asymptotic relation (6.19). This estimator depends only on the counts of new infected individuals and is easy to compute as follows~:

First, select an autoregressive order $H$ and then regress $N_{12} (t)$ on $N_{12} (t-1), \ldots, N_{12} (t-H)$ [without intercept] by OLS, for $t=H+1, \ldots, T$. If $\hat{\gamma} (s), s=1,\ldots,H$, are the estimated regression coefficients, define the estimator of the reproductive number as~:

\begin{equation}
  \hat{R}^{AR}_{00} = \Sum^H_{s=1} \hat{\gamma} (s).
\end{equation}

This estimator has a variance that will increase with $H$, since more underlying parameters have to be estimated. It has also the drawback of being computable only after at least $2H+1$ days, due to the lag and the minimal number of observations necessary to identify the autoregressive parameter. These estimates have been computed for $H=7, 14, 21$ days in a non-rolling way on the same set of simulated data used to generate Figure 6.\vspace{1em}

\newpage
\textbf{Figure 7 : The Autoregression Estimate}

\begin{flushleft}
\includegraphics[width=0.9\textwidth,height=0.9\textheight,keepaspectratio]{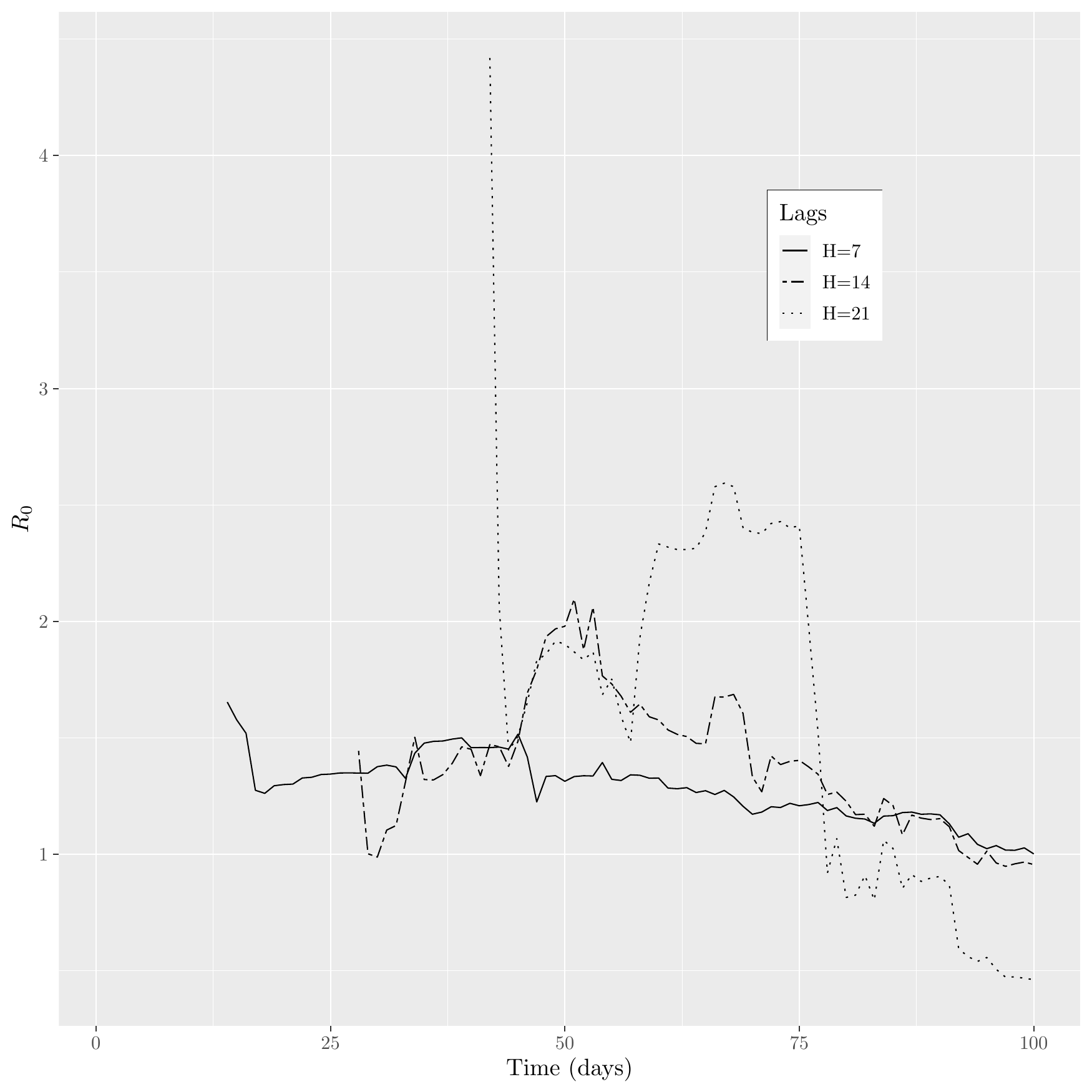}
\end{flushleft}

The variability effect and the impossibility to use it at the beginning of the disease are clearly observed. We also note that these estimators do not converge to the true value. Indeed this approach is also subject to the causality and linearization biases. The bias is observed in Figure 7 with an underestimation of $R_0$.

\section{Concluding Remarks}

The estimated reproductive numbers are used as basic tools to follow the progression of an epidemic such as COVID-19 and monitor the (changes in) health policy. For instance, specific partial lockdown policies may be introduced if the estimated $R_0$ is larger than 1.5. Such policies neglect the variability of both this notion and its approximations (estimates). We have considered this question in the framework of a time discretized SIR model and shown that this variability can be due to the definition itself which is time dependent, sometime author dependent, or to an omitted underlying heterogeneity. It is also a consequence of the different estimation methods that are used, with bias and uncertainty that depend on the available information.

As a by-product we have shown that the estimate of $R_0$ based on the Poisson approximate likelihood of the SIR model, used in a rolling way, with possibly a prior on parameters (see Appendix 4 for the Bayesian estimation) is as simple as the approach suggested in the standard  EpiEstim, with two advantages: it is using the information of both new infected and currently infected people, and does not fix arbitrarily the infectiousness profile.

As mentioned in the text, Thompson et al. (2019) highlight issues related to the use of the standard EpiEstim and propose an improved version of the package to correct some of the drawbacks of the standard one. They propose to use in real time two series of data: the counts of new infected people and ``up-to-date observations of serial intervals". In this approach there will be (as in the rolling approach based on SIR) a larger information set, and this updating will also introduce path varying mean and standard deviation of the distribution of the serial interval. With this improved version, the two approaches will not differ greatly by the underlying model on which they are based, (see the discussion in 6.4.1, 6.4.2) but instead by the observations they are using to calibrate the parameters. That is, the counts of new infected and currently infected people in the SIR model based estimator and the counts of new infected and data on pairs of infector/infectee obtained by tracing.

The choice of approach should largely depend on the availability (and cost) of such data, especially at the beginning of the epidemic, and on their reliability. In particular, the available data are currently incomplete, since they do not account for the undetected asymptomatic people [see Gourieroux, Jasiak (2020 a], and they are left and right censored for tracing of the pairs infector/infectee.

The SIR model has been chosen since the different estimation approaches were implicitly based on this model and it has facilitated the discussions and comparisons. Clearly, to obtain a more complete picture, similar exercises would have to be done on models with more features. The aim of this extended analysis would be to account for the difference between the infection and infectious period and also incorporate a stochastically time varying contagion parameter component [see e.g. Gourieroux, Lu (2020)].

\newpage
\begin{center}
    {\bf Appendix 1 }\vspace{1em}

    \setcounter{equation}{0}\def\theequation{a.\arabic{equation}}

\textbf{The Continuous Time SIR Model}
\end{center}\vspace{1em}

The SIR model is usually written as a continuous time deterministic model. The standard notations are~:

$$
x(t) = p_1 (t), y(t) = p_2 (t), z(t) = p_3 (t).
$$

This model defines the dynamics of the cross-sectional structure by the system of differential equations~:

\begin{equation}
  \left\{ \begin{array}{lcl}
        \Frac{dx(t)}{dt} & = & -\alpha x(t) y(t), \\ \\
        \Frac{dy (t)}{dt} & = & \alpha x(t) y (t) - \gamma y (t), \\ \\
         \Frac{dz (t)}{dt} & = & \gamma y (t),
          \end{array}
  \right.
\end{equation}

\noindent where $\alpha, \gamma$ are positive parameters.\vspace{1em}

This differential system admits a closed form solution derived rather recently [Harko, Lobo, Mak (2014), Section 2, eq (17)-(20)]. This solution depends on parameters $\alpha, \gamma$ and on starting values $x(0), y(0), z(0).$\vspace{1em}

Let us consider the integral equations~:

\begin{eqnarray}
  t& =& \Int^{u(t)}_{\exp [-\Frac{\alpha}{\gamma} z(0)]} \Frac{dv}{v[-\alpha-\gamma \log v + \alpha x(0) \exp [\Frac{\alpha}{\gamma} z(0] v]}, \\
  &\equiv & G [u(t); \alpha, \gamma, p(0)], \nonumber
\end{eqnarray}

\noindent where $p(0) = [x(0), y(0), z(0)]'$ is the initial structure. Then the solution is~:

$$
\left\{
\begin{array}{lcl}
  x(t) & = & x(0) \exp [\Frac{\alpha}{\gamma} z(0)] G^{-1} [t; \alpha, \gamma, p(0)], \\ \\
  y(t) & = & 1-x(t) - z(t), \\ \\
  z(t) & = & -\Frac{\alpha}{\gamma} G^{-1} [t;\beta, \gamma, p(0)].
\end{array}
\right.
$$

The knowledge of the solution allows to derive the following results.\vspace{1em}

i) $x(t)$ decreases to a limiting value $x(\infty)$.

ii) $z(t)$ increases to a limiting value $z(\infty)$.

iii) $y(t)$ usually increases to a peak, then decreases to $y(\infty) = 0$.

iv) There is herd immunity that is $x(\infty) > 0$, and this final size is equal to the solution of~:

$$
x(\infty) - x(0) - y(0) - \Frac{c}{a} \log [x(\infty)/x(0)] = 0.
$$

v) The herd immunity can be reached in a finite time.\vspace{1em}

[see e.g. Kermack, McKendrick (1927), Hethcote (2000), or Ma, Earn (2006) for the expression of the final size, and Gourieroux, Lu (2020), for property v)].\vspace{1em}

An analogue discrete time deterministic model is~:

$$
\left\{
\begin{array}{lcl}
x(t) & = & x(t-1) - a x(t-1) y (t-1), \\ \\
y(t) & = & y(t-1) + a x(t-1) y (t-1)-c y (t-1), \\ \\
z(t) & = & z(t-1) + cy (t-1).
\end{array}
\right.
$$

This analogue is not the exact time discretized continuous time SIR. In particular the parameters $a,c$ have interpretations that slightly differ from $\alpha, \gamma$, and may depend on the time-step of the discretization. Moreover, in nonlinear dynamic systems, such a Euler discretization might change the dynamic properties of the trajectories. However, it is known that properties i), ii), iii) of the trajectories are still satisfied and that there is always herd immunity [Allen (1994)]. However, the herd immunity cannot be reached in a finite time and the expression of the final size is not known under closed form.

This discrete time analogue is exactly the mechanistic model derived in Section 3.2.\vspace{1em}

\begin{center}
    {\bf Appendix 2 }\vspace{1em}

\textbf{Statistical Inference}
\end{center}\vspace{1em}

\textbf{1. Likelihood function.}\vspace{1em}

The individual histories are equivalently characterized by the sequence of dummy variables~:

$$
\begin{array}{lcl}
z_{jit} &=& 1,\, \mbox{if individual}\; i\;\mbox{is in state}\; j\; \mbox{at date}\; t,\\ \\
&&0,\; \mbox{otherwise.}
\end{array}
$$

Then, by applying the Bayes formula, the likelihood is equal to~:\footnote{with the appropriate convention for treating the absorbing state.}

$$
\begin{array}{lcl}
  l(a,c) & = & \Pi^2_{j=1} \Pi^3_{k=1} \Pi^n_{i=1} \Pi^T_{t=1} \left[p_{jk}(t;a,c)^{ z_{ij,(t-1)} z_{ikt}}\right]\\ \\
         & = & \Pi^2_{j=1} \Pi^3_{k=1} \left[ p_{kj} (t;a,c)^{ \Sum^n_{i=1} \Sum^T_{t=1} z_{ij (t-1)} z_{ikt}}\right] \\ \\
         & = & \Pi^2_{j=1} \Pi^3_{k=1} p_{jk} (t;a,c)^{ \Sum^T_{t=1} N_{jk} (t)},
\end{array}
$$

\noindent where the transition probabilities may depend on $N_2 (t-1).$ This explains why the $N_{jk} (t), j,k=1,2,3$ define a sufficient statistics.\vspace{1em}

\textbf{2. Decomposition of the log-likelihood function}\vspace{1em}

We deduce~:

$$
\begin{array}{lcl}
  L(a,c) & = & \log l(a,c) \\ \\
  &=& \Sum^3_{k=1} \left[ \Sum^T_{t=1} N_{1k} (t) \log p_{1k} (t,a)\right] \\ \\
  &+& \Sum^3_{k=1} \left[ \Sum^T_{t=1} N_{2k} (t) \log p_{2k} (t,c)\right] \\ \\
  &\equiv & L_1 (a) + L_2 (c),
\end{array}
$$

\noindent noting that the transition probabilities of the first row (resp. the second row) depend on $a$ [resp. $c$] only.
\newpage

\begin{center}
\textbf{Appendix 3}
\end{center}\vspace{1em}

\textbf{Finite Sample Properties of the Unfeasible Gaussian ML Estimators}\vspace{1em}

We provide for the unfeasible Gamma ML estimator the Figures a.1, a.2, that are the analogues of Figures 4, 5 given in the text for the approximate Poisson ML estimator. These distributions are similar to the distributions for the Poisson . Nevertheless Figure 6 shows that their evolutions with the number of observations are highly different.\vspace{1em}
\newpage

\textbf{Figure a.1 : Distribution of Approximate Unfeasible Gaussian Estimators, $N_2 (0) = 100$}\vspace{1em}

\makebox[\textwidth]{
\begin{tabular}{cc}
  \includegraphics[width=80mm]{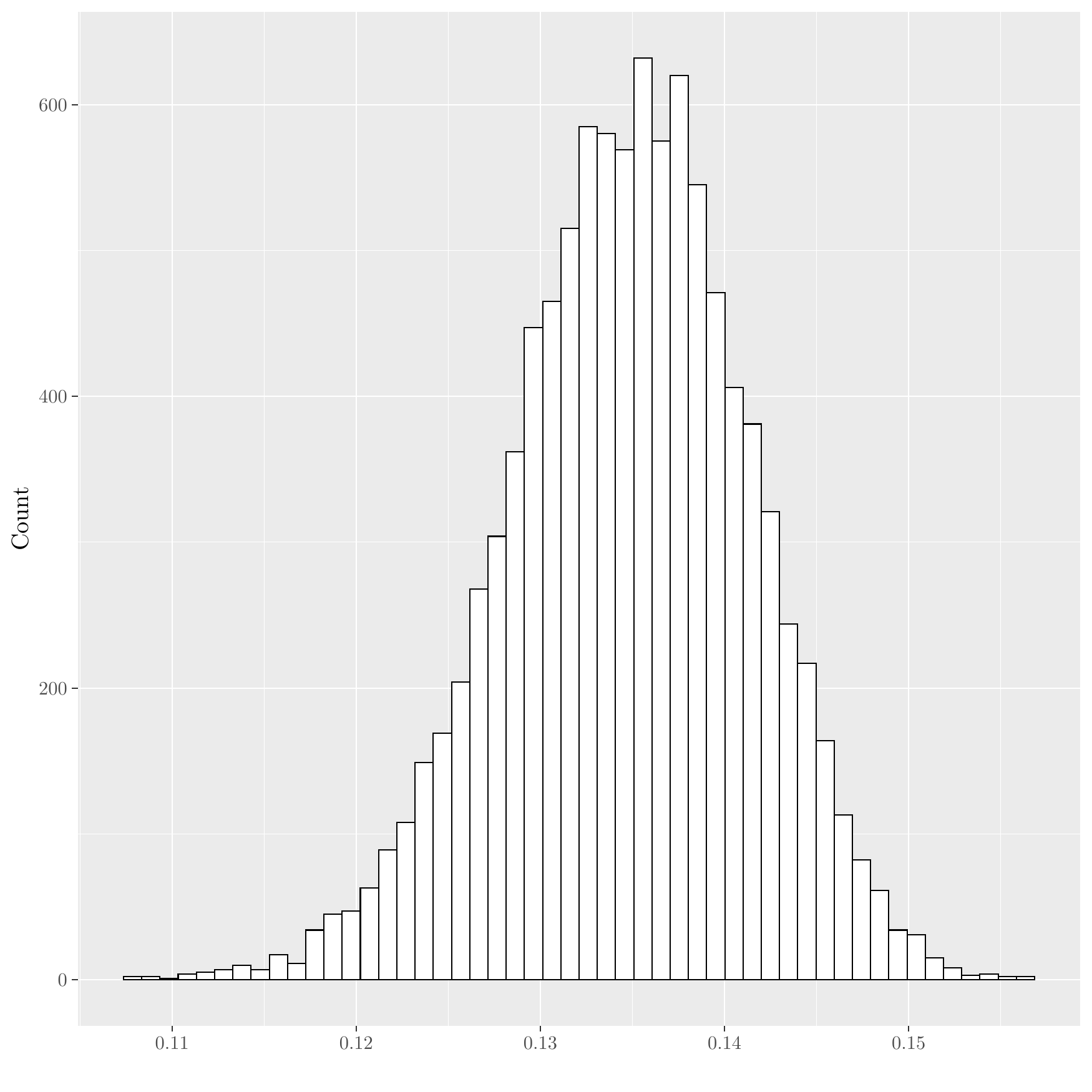} &   \includegraphics[width=80mm]{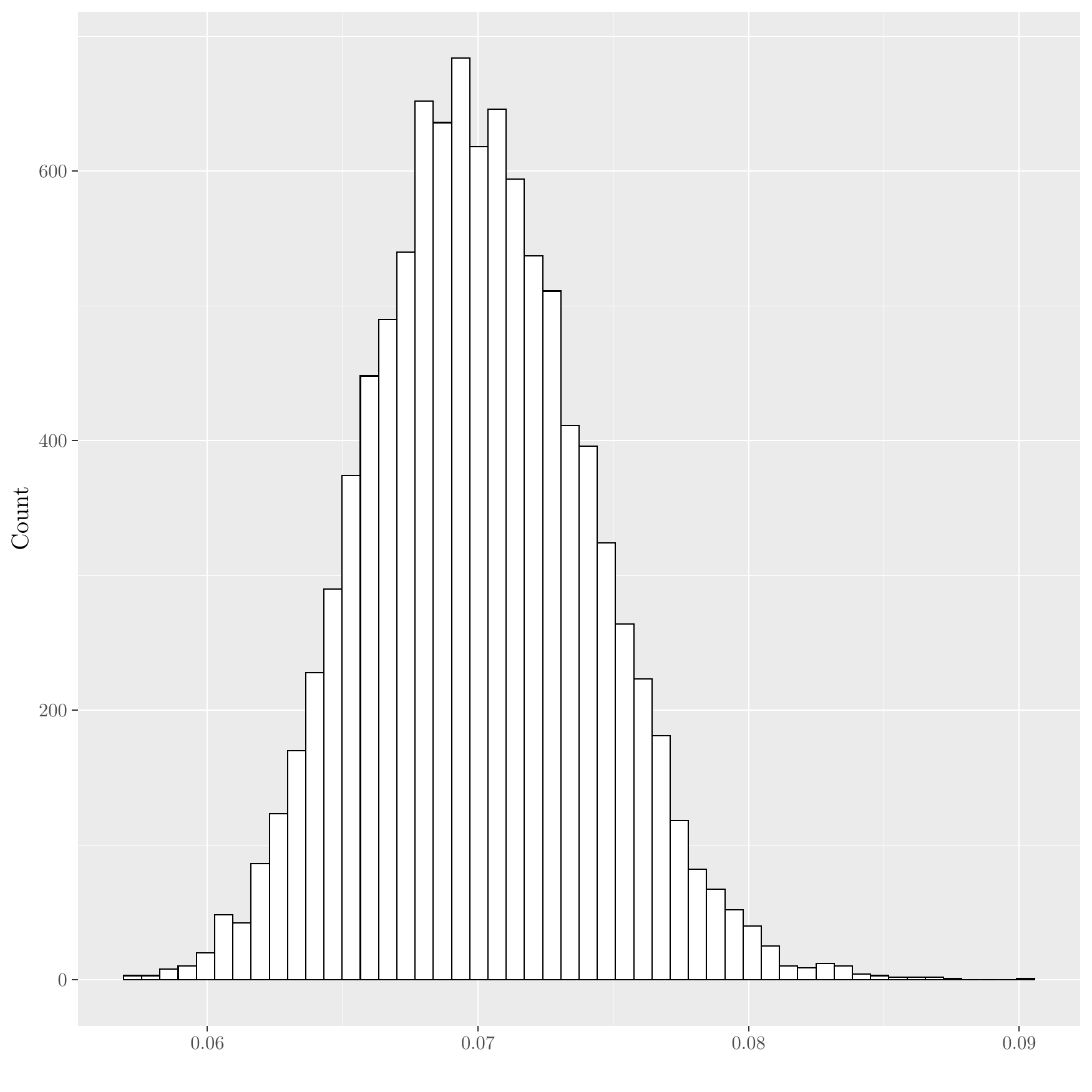} \\
$\hat{a}$ & $\hat{c}$ \\[6pt]
\multicolumn{2}{c}{\includegraphics[width=80mm]{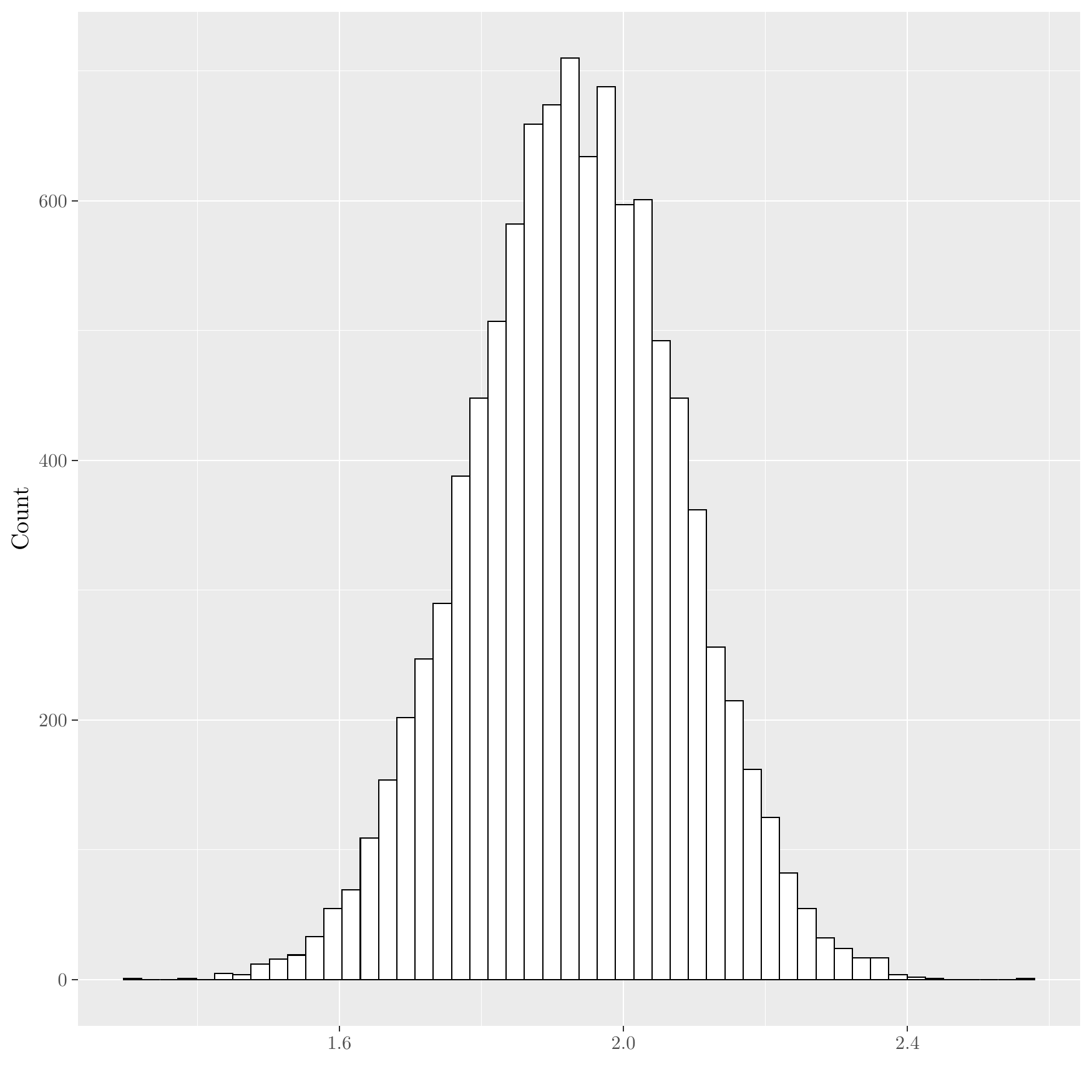} }\\
\multicolumn{2}{c}{$\hat{R}_0$}
\end{tabular}} 
\newpage

\textbf{Figure a.2 : Distribution of Approximate Unfeasible Gaussian Estimators, $N_2 (0) = 1000$}\vspace{1em}

\makebox[\textwidth]{
\begin{tabular}{cc}
  \includegraphics[width=80mm]{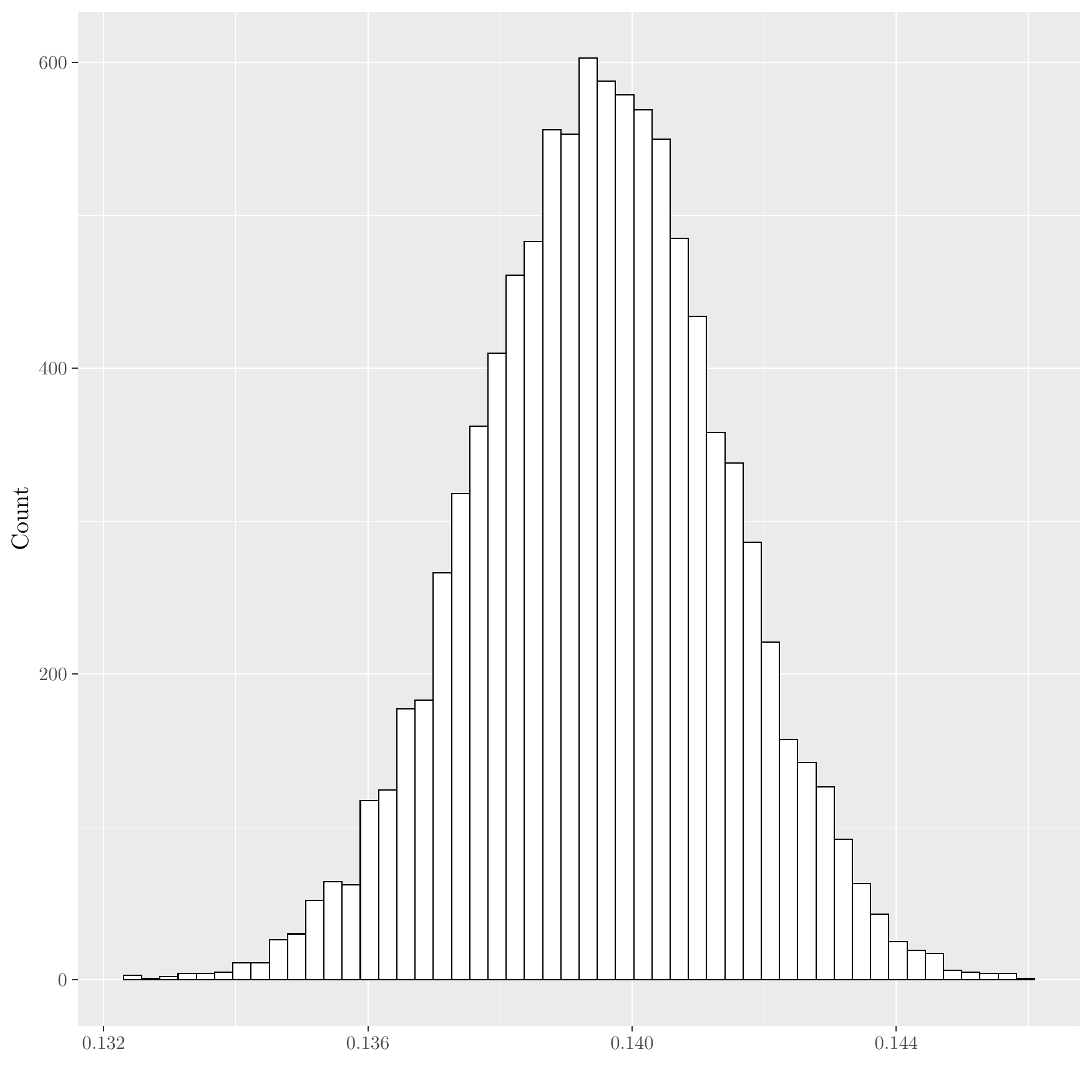} &   \includegraphics[width=80mm]{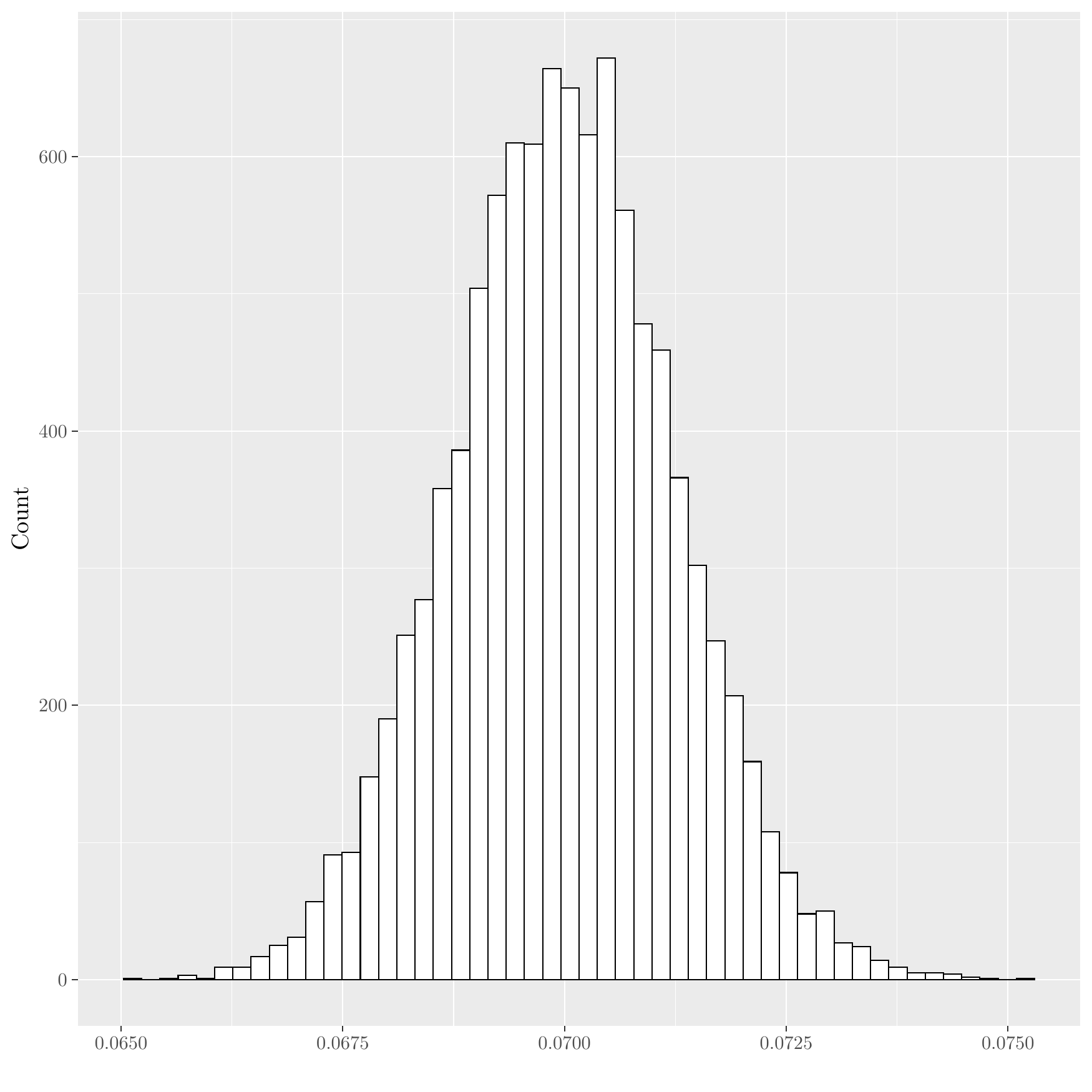} \\
$\hat{a}$ & $\hat{c}$ \\[6pt]
\multicolumn{2}{c}{\includegraphics[width=80mm]{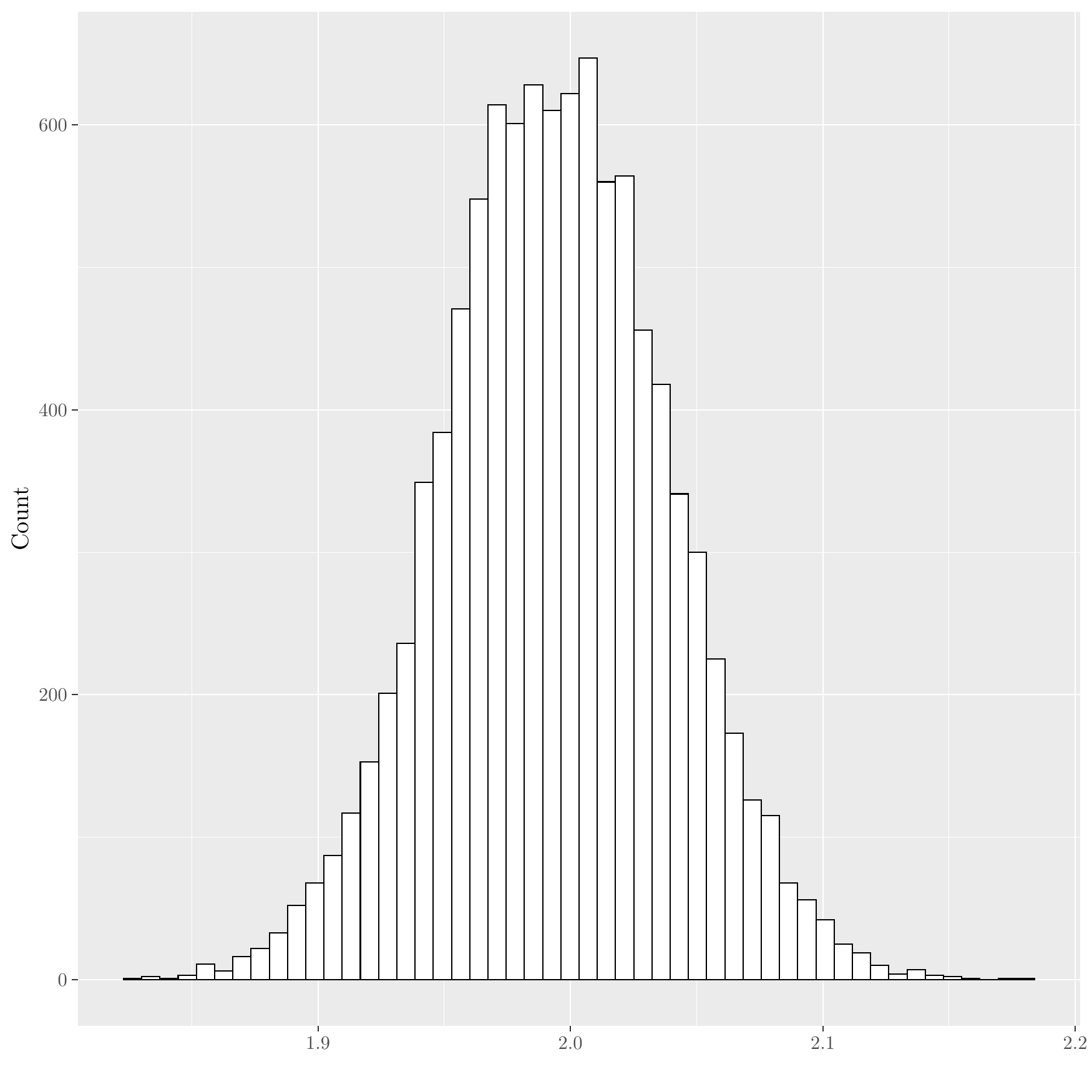} }\\
\multicolumn{2}{c}{$\hat{R}_0$}
\end{tabular} }

\newpage

\begin{center}
    {\bf Appendix 4 }\vspace{1em}

\textbf{Bayesian Estimators}
\end{center}\vspace{1em}

Several authors have considered Bayesian estimation approaches [see e.g. Cori et al. (2019), Webappendix 1]. To facilitate the comparisons, we consider below Bayesian estimation approaches for the Poisson approximate likelihood. As noted in the literature the Poisson likelihood has an expression in parameters $a,c$, that allows for a conjugate prior for these parameters. More precisely, the following result is easily derived.\vspace{1em}

\textbf{Proposition :} For the Poisson approximate likelihood,

i) A conjugate prior for $a,c$ is such that $a$ and $c$ are independent with gamma distributions $\gamma (\nu_a, \lambda_a)$ and $\gamma (\nu_c, \lambda_c)$, respectively.

ii) For this prior the posterior is such that~: $a$ and $c$ are independent with gamma distributions~:

$$
\begin{array}{l}
\gamma [\Sum^T_{t=1} N_{12} (t) +
 \nu_a, \Sum^T_{t=1} [N_1 (t-1) \hat{p}_2 (t-1)] + \lambda_a],\\ \\
 \gamma [\Sum^T_{t=1} N_{23} (t) +
 \nu_c, \Sum^T_{t=1} [N_2 (t-1) +  \lambda_c],
 \end{array}
$$

\noindent respectively.\vspace{1em}

iii) Let us denote $\nu_a (t), \nu_c (t), \lambda_a (t), \lambda_c (t)$ the degrees of freedom and scales of the posterior distributions of $a$ and $c$. Then the posterior distribution of $R_0 = a/c$ is such that: $\Frac{\lambda_a (t) \nu_c (t)}{\lambda_c (t) \nu_a (t)} R_0$ follows a Fisher distribution $F (2 \nu_a (t), 2 \nu_c (t))$.\vspace{1em}

This Bayesian analysis differs from the derivation in Cori et al. (2013). Indeed the conjugate prior is naturally introduced on parameters $a$ and $c$ by gamma distributions, whereas they introduce a nonconjugate prior on $R^i_0$ and the infection profile $w (s)$, characterized by $c$ in the SIR framework. This modifies significantly the posterior of $R_0$. The posterior mean of $R_0$ is : $\Frac{\lambda_c (t) \nu_a (t)}{\lambda_a (t)} \Frac{\nu_c (t)}{\nu_c (t) -1}$, if $\nu_c (t) > 1$, and does not exist, otherwise.

The reason for the non existence of the posterior mean is similar to the reason for the nonexistence of the Poisson approximate maximum likelihood estimator. If we observe no recovery, or if there is a prior for a long infectious period, the ML or Bayesian approaches can provide posterior distributions with fat tail.

\end{document}